\documentclass[referee]{raa}           
\usepackage{graphicx,times}
\usepackage{natbib}
\usepackage{amssymb,amsmath}
\usepackage{graphicx}
\usepackage{subcaption}
\usepackage{placeins}
\bibpunct{(}{)}{;}{a}{}{,}

\usepackage[a4paper=true,pagebackref=true]{hyperref}
\hypersetup{pdftitle = The title of my PDF, pdfauthor = My name, pdfsubject= The subject, pdfkeywords = keyword1 keyword2 keyword3} 
\hypersetup{colorlinks = true, linkcolor = green, anchorcolor = red, citecolor = blue, filecolor = red, pagecolor = red, urlcolor = red}

\begin{document}

   \title{PHYSICAL PROPERTIES OF THREE SHORT PERIOD CLOSE BINARIES: KIC 2715417, KIC 6050116 AND KIC 6287172 
$^*$
\footnotetext{\small $*$ Discovered by Kepler Space Telescope}
}

 \volnopage{ {\bf 2018} Vol.\ {\bf 18} No. {\bf 8}, 000--000}
   \setcounter{page}{1}

   \author{M. A. NegmEldin\inst{1}, A. Essam\inst{2}, Shahinaz M. Yousef\inst{1}
   }

   \institute{ Astronomy,Space Science and Meteorology Department, Faculty of Science, Cairo University, Cairo, 12613, Egypt; {\it Negmeldeen@cu.edu.eg}\\
        \and
             Department of Astronomy, National Research Institute of Astronomy and Geophysics, Helwan, 11421, Egypt\\
\vs \no
   {\small Received 2018 January 11; accepted 2018 June 13}
}

\abstract{We present the physical parameters of three short period close binaries using the observed data from Kepler Space Telescope. All of these observations were taken in a single bandpass (which approximates the Johnson $V$-band). Our three systems are KIC 2715417, KIC 6050116 and KIC 6287172. The first system KIC 2715417 is considered a semi-detached system with the secondary component filling the Roche lobe.  The second system KIC 6050116 is an overcontact system, while the third system KIC 6287172 belongs to ellipsoidal variables (ELV) as deduced from the Roche lobe geometry. For photometric analysis, we used PHOEBE software package which is based on the Wilson \& Devinney code. Due to lack of spectroscopic data, the photometric mass ratios are determined from the analyses of the light curves using the $Q$-search method.  The absolute parameters are determined using three different methods (Harmanec, Maceroni \& Van'tVeer and Gazeas \& Niarchos).
}

   \authorrunning{$M.~A.~NegmEldin~et~al.$:~Physical properties of three short period close binaries }            
   \titlerunning{$M.~A.~NegmEldin~et~al.$:~Physical properties of three short period close binaries }  
   \maketitle

%
\section{Introduction}           
\label{sect:intro}

We have many types of close binary systems as a result of close binary evolution. These types can be classified as follows:

The type of Semi-Detached systems, with the surfaces of the low mass components, are found in contact with the inner Lagrangian surfaces (the critical potential surfaces) and the surfaces of the more massive components are bounded within a separate equipotential surface.
The Beta Lyrae systems, where one of the two components has become a giant or supergiant in the course of its evolution, the matter can be escaped from the critical potential surface at which the gravitation at their surface is weak to transfer into the other component. 
W UMa type or late-type of contact systems have both of their components filling the inner Lagrangian surfaces and share the common envelope. Ellipsoidal variables can be described as very close binaries with the two components are non-spherical shapes (ellipsoidal shapes) due to the mutual gravitation.
The binary systems can be considered as close binary for the following reasons: 

  The close binaries have short periods. The distance separating the two components is comparable to their size. The two components are close enough that their shapes are distorted by mutual gravitation forces to non-spherical shapes such as in the type of ellipsoidal variables.
  The surface of both components overflow their critical Lagrangian surfaces and share a common envelope such as in the case of overcontact type. When one of its two components become giant or supergiant in the final stages of its evolution, matter may freely flow from one component to the other as in the case of Beta Lyrae type.

\subsection{OBSERVATIONS}
\label{sect:Obs}

Kepler is a space telescope launched by NASA on March 7, 2009 to discover extra solar planets orbiting around other stars in the field of Cygnus constellation. Kepler has a primary mirror equal to 1.4 meters in diameter. The FOV of Kepler spacecraft has a 105 $deg^2$. The photometer is composed of an array of 42 CCDs, each CCD has 2200x1024 pixels\footnote{http://www.nasa.gov/missionpages/kepler/spacecraft/index.html}. The Kepler’s catalogue ID, coordinates, V magnitude, color excess and the interstellar extinction of the target objects are listed in Table~\ref{tab1}

\begin{table}[h]
\bc
\begin{minipage}[]{100mm}
\caption[]{Catalogue ID, Coordinates, and V Magnitude\label{tab1}}\end{minipage}
\setlength{\tabcolsep}{1pt}
\small
 \begin{tabular}{lccccc}
  \hline\noalign{\smallskip}
Kepler$'$s Catalogue ID~~~~~&~~~~$\alpha$2000~~~~~ &~~~~ $\delta$2000~~~~~~&~~~~~~$V${\tiny mag}~~~~~~&~~~~~E(B-V)~~~~~~&~~~~~$A_{\nu}$~~~~~~ \\
 & {\tiny {$(hh:mm:ss.ss)$}}~~~~&~~~~~~{\tiny {$(\pm dd:mm:ss.ss)$}}&&&\\
  \hline\noalign{\smallskip}
KIC 2715417&19:27:52.565&+37:55:39.97&14.070&0.063&0.427\\
KIC 6050116&19:36:33.113&+41:20:22.56&14.258&0.059&0.369\\
KIC 6287172&19:29:59.69&+41:37:45.0&12.714& 0.114&0.339\\

  \noalign{\smallskip}\hline
\end{tabular}
\ec
\end{table}

\begin{figure}[h]
        \centering
        \includegraphics[width=8.0cm, angle=0]{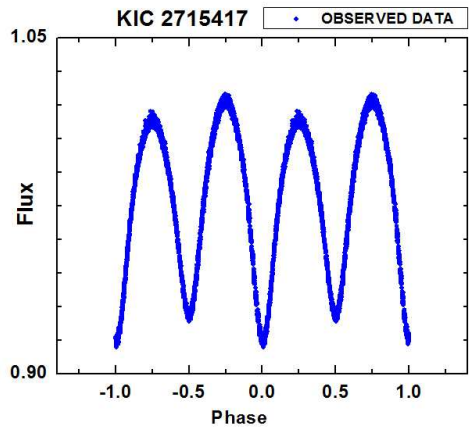}
        \begin{minipage}[]{85mm}
        \caption{Light Curve of the system KIC 2715417\label{Fig1}.}
        \end{minipage} 
\end{figure}

\begin{figure}[h]
        \centering
         \includegraphics[width=8.0cm, angle=0]{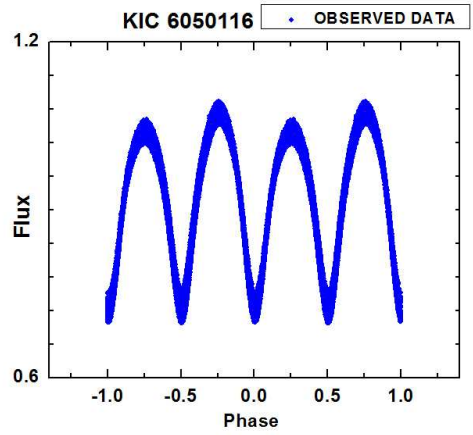}
        \begin{minipage}[]{85mm}
        \caption{Light Curve of the system KIC 6050116\label{Fig2}. }
        \end{minipage} 
\end{figure}

\begin{figure}[h]
        \centering
         \includegraphics[width=8.0cm, angle=0]{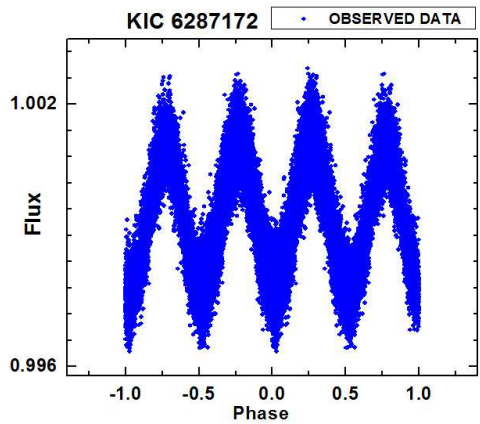}
         \begin{minipage}[]{85mm}
         \caption{Light Curve of the system KIC 6287172\label{Fig3}.}
          \end{minipage} 
\end{figure}

       From the Kepler data archive\footnote{http://keplerebs.villanova.edu/}, we have collected the observations of the light curves for the above three systems. Observations of the KIC 2715417 system started at 2454964.51259 JD and ended at 2456390.95883 JD. The light curve of these data is shown in Figure~\ref{Fig1}.  In the case of KIC 6050116 system, observations started at 2454964.51229 JD and ended at 2456390.95881 JD and the corresponding light curve is shown in Figure ~\ref{Fig2}. For the system KIC 6287172, the observations started at 2454953.53905 JD and ended at 2456390.95893 JD. The corresponding light curve of this system is shown in Figure~\ref{Fig3}.\\

 \FloatBarrier  
 
   In the case of the eclipsing binary system KIC 2715417 and from the Kepler Input Catalog\footnote{http://archive.stsci.edu/kepler/keplerfov/search.php}, we found that the effective temperature of the primary component is $5189^{\circ}$ K, the Kepler magnitude is 14.070{\tiny mag}, metallicity = -0.458, $log_{10}$ G ($log_{10}$ Surface gravity) = 4.830 and the color Excess reddening $E(B-V)$= 0.063. It is considered a semi-detached binary with the secondary component filling the Roche lobe. The represented ephemeris of the system can be written as: 
\begin{equation}
 HJD(Min I)=2454964.666158  + 0.2364399 \times {E} 
\label{eq1}
\end{equation}
Where HJD (Min I) represents the minima epoch times in Heliocentric Julian Date and E is the integer number of cycles.

      For the eclipsing binary system KIC 6050116 and from the Kepler Input Catalog, it is found that the effective temperature of the primary component is $4569^{\circ}$ K, the Kepler magnitude equals 14.25{\tiny mag}, Metallicity (Solar= 0.0134) =-0.849, $log_{10}$ G ($Log_{10}$ surface gravity) = 4.524 and color Excess reddening $E(B-V)$= 0.059, it is an overcontact system. 
The represented ephemeris of the system can be written as:
\begin{equation}
 HJD(Min I)=2454964.708006   + 0.2399081  \times {E} 
 \label{eq2}
\end{equation}
    Using the Same Kepler Input Catalog for the eclipsing binary system KIC 6287172, we also found that the effective temperature of the primary component is $6646^{\circ}$ K and the Kepler magnitude is equal to 12.714{\tiny mag} and Metallicity (Solar=0.0134) = -0.470 and $log_{10}$ G = 4.286 and color Excess reddening $E(B-V)$= 0.114, it is considered an ellipsoidal variable ($ELV$).
The represented ephemeris of the system can be written as:
\begin{equation}
HJD(Min I)=2454953.651911  + 0.2038732 \times {E}
\label{eq3}
\end{equation}

\section{LIGHT CURVE ANALYSES}
\label{sect:LCs}
      For the three systems (KIC 2715417, KIC 6050116, and KIC 6287172) it appears that their temperatures are below $7200^{\circ}$ K which indicates that they are convective, hence the gravity darkening for the convective stars $g_{1}=g_{2}=0.32$ \citep{Lucy+1967}.  The bolometric albedo $A_{1}=A_{2}=0.5$ \citep{Rucinski+1969} and the limb darkening coefficients were adopted from \cite{VanHamme+1993} based on the linear cosine law model.
      
\begin{figure}[h]
  \centering
 \includegraphics[width=8.0cm, angle=0]{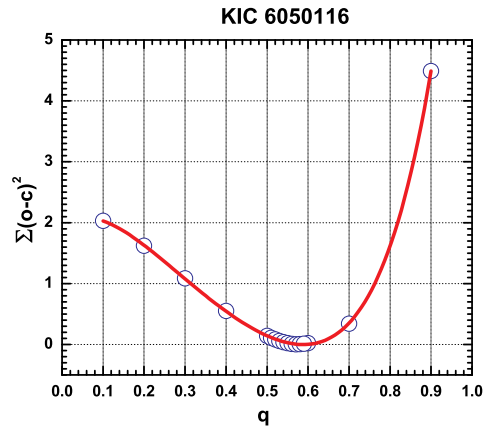}
 \begin{minipage}[]{85mm}
\caption{The $Q$-search diagram for the system KIC 2715417\label{Fig4}. } 
\end{minipage}
\end{figure}

\begin{figure}[h]
  \centering
 \includegraphics[width=8.0cm, angle=0]{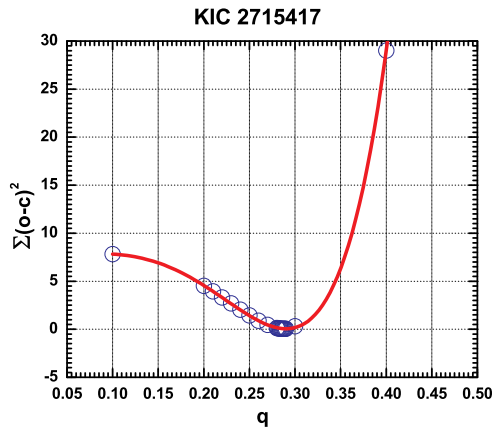}
 \begin{minipage}[]{85mm}
\caption{The $Q$-search diagram for the system KIC 6050116\label{Fig5}. } 
\end{minipage}
\end{figure}

\begin{figure}[h] 
  \centering
 \includegraphics[width=8.0cm, angle=0]{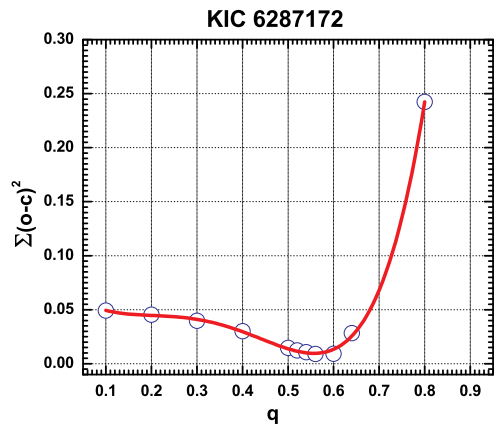}
 \begin{minipage}[]{85mm}
\caption{The $Q$-search diagram for the system KIC 6287172\label{Fig6}. } 
\end{minipage}
\end{figure}

    With the lacking of any spectroscopic data for these three systems, it is difficult to determine the mass ratio accurately. Thus, we used the $Q$-search method as discussed in \cite{Deb+etal+2010} to determine the approximate value of the photometric mass ratios for these systems. 
   We have adjusted the PHOEBE software model of unconstrained binary system to estimate a set of parameters that represent the observed light curve. In order to determine the photometric mass ratio, we used a series of mass ratios values ranging from 0.1 to 1.0 in steps of 0.1 after that we took the three values of mass ratio corresponding to the minimum values of the residuals. Then, we used the same technique but with steps of 0.01. 
    For each value of mass ratio, we obtained the sum of the squared deviations (residuals) from the fitting solutions of the light curve modeling. Figure~\ref{Fig4} shows that the minimum occurs at value q= 0.29 for the system KIC 2715417. Figure~\ref{Fig5} shows that the minimum occurs at value q=0.57 for the system KIC 6050116. Finally Figure~\ref{Fig6} shows that the minimum occurs at value q=0.61 for the system KIC 6287172 which is not to be trusted due to very small inclination of (ELV) Type. Rather, the models offered here are the best that can be done with the available data.

 \FloatBarrier
 
\subsection{Light curve analyses of the system kic 2715417}
\label{sect:KIC 2715417}

The light curve of the system KIC 2715417 in the $V$-band as shown in Figure~\ref{Fig1} has been analyzed using the PHOEBE package, version 0.31a \citep{Prsa+Zwitter+2005} which is based on the code of \cite{Wilson+Devinney+1971} . From the Kepler Input Catalogue, it is found that the corresponding temperature $T_{1}= 5189^{\circ}$ K. First, from the PHOEBE software we have used the model of unconstrained binary system to find approximate values of the parameters that represent the observed light curve.
     The data obtained from the PHOEBE model of unconstrained binary system, were analyzed to find the fillout factor (f) using the software Binary Maker 3 (BM3). This software is a modification of the parameter defined by \cite{Lucy+Wilson+1979} to specify equipotentials for contact, overcontact and detached systems after specifying the mass ratio. In the case of detached type, the fill-out is given by formula (A).  For overcontact, the fill-out is given by formula (B):\\

 $~~~~~~~~~~ {f}=\frac {\Omega_{inner}}{\Omega}-1, ~~~~~~~~~~~for~~~~~~~~{\Omega_{inner}} < {\Omega}~~~(Detached)~~~~~~~~(A)$\\

 $~~~~~~~~~~{f}=\frac {\Omega_{inner} - \Omega}{\Omega_{inner} - \Omega_{outer}}, ~~~~for~~~~~~~~{\Omega_{inner}} \geqslant {\Omega}~~~(Overcontact)~~~~~~(B)$\\

Thus, the fill-out factor (f) for detached stars will lie between $(-1 < f < 0)$. The fill-out factor (f) for overcontact systems will lie between $(0 < f < 1)$. In the case of the contact system, the fill-out factors of two stars equals zero $(f = 0)$. When (f) is near zero, the system can be described as a near contact system (e.g. the smaller star has a fill-out factor $(f_{2})= 0.00$ and the larger star has $(f_{1}) = -0.02$ ).

     In our case, we found the fill-out factor of the primary component $f_{1}= - 0.1150$, and the secondary component has fill-out factor $f_{2}= 0.0013$.  According to the above mentioned rules, the best model describing the binary system KIC 2715417 is a near contact system. The best photometric fitting has been reached after several runs, which shows that the primary component is massive and hotter than the secondary one, with a temperature difference of about $478^{\circ}$ K.
      
\begin{figure}[h] 
  \centering
 \includegraphics[width=8.0cm, angle=0]{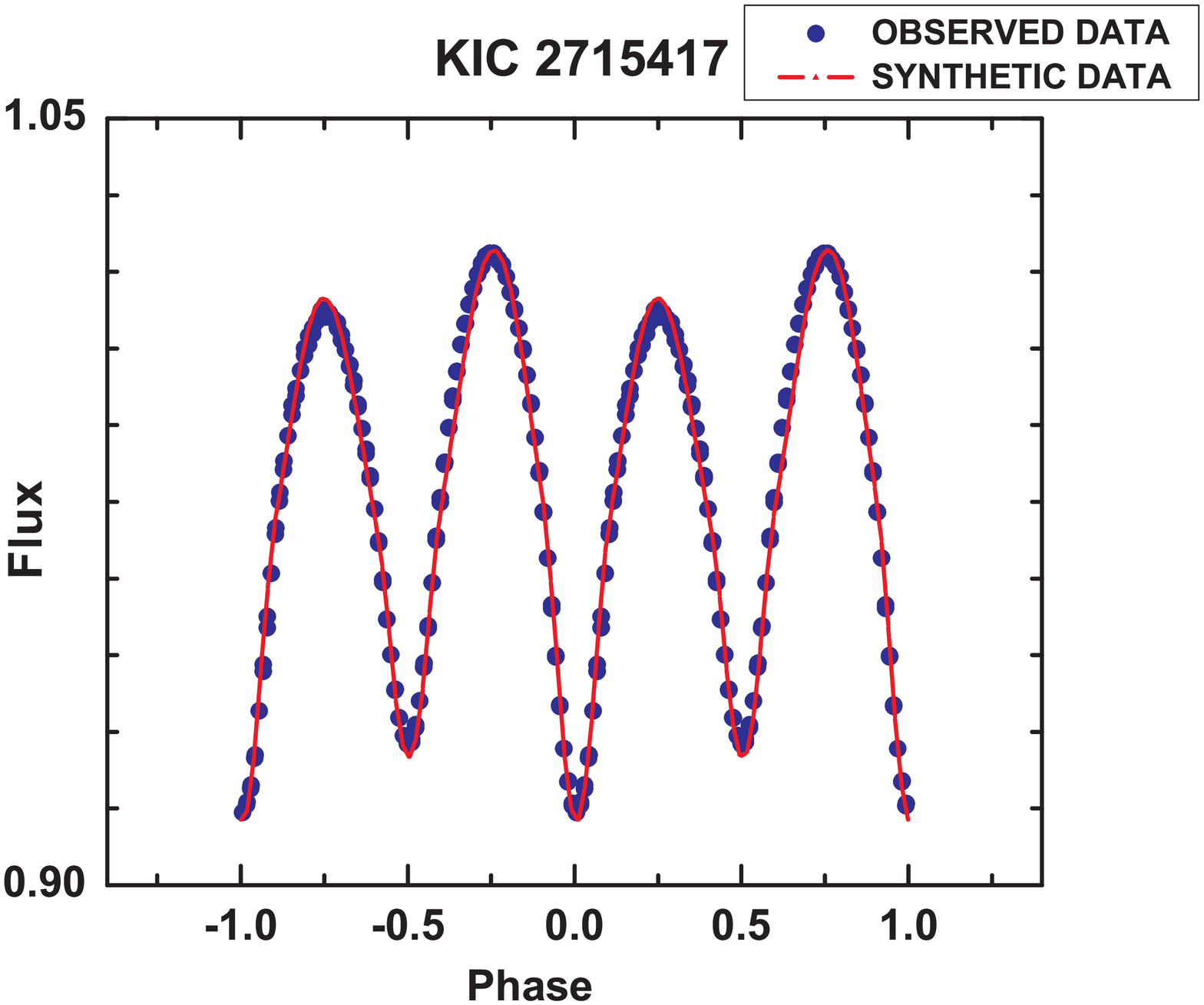}
 \begin{minipage}[]{85mm}
\caption{Light curve in $V$ filter showing the coherence between observed and synthetic curves\label{Fig7}. } 
\end{minipage}
\end{figure}

\begin{figure}[h]
  \centering
 \includegraphics[width=8.0cm, angle=0]{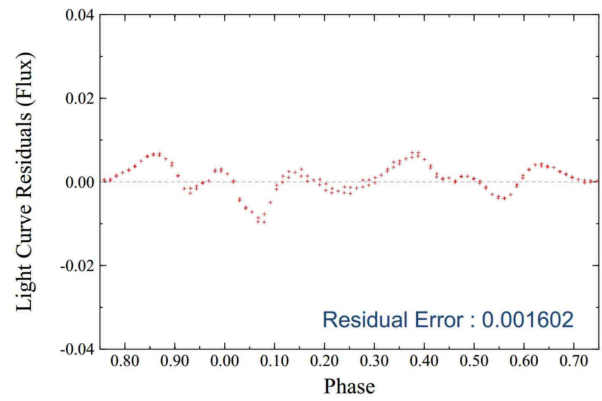}
 \begin{minipage}[]{85mm}
\caption{Light curve residual error for different phases for the system KIC 2715417\label{Fig8}. } 
\end{minipage}
\end{figure}

      The orbital and physical parameters of the system KIC 2715417 are listed in Table~\ref{tab2}. Figure~\ref{Fig7} displays the observed light curve for the interval 2454964.51259 JD - 2454972.99293 JD, together with the synthetic curve in the $V$-band while Figure~\ref{Fig8} displays the light curve residual error for different phases. According to the effective temperature of both the primary and secondary components of the system KIC 2715417 and from the calibration of Morgan-Keenan (MK) spectral types for the main sequence class\citep{Drilling+Landolt+2002}, the spectral types are nearest to $K_{0}$ and $K_{2}$ respectively. 
      
\begin{figure}[h]
\begin{subfigure}{0.5\textwidth}
\includegraphics[width=6.0cm, angle=0]{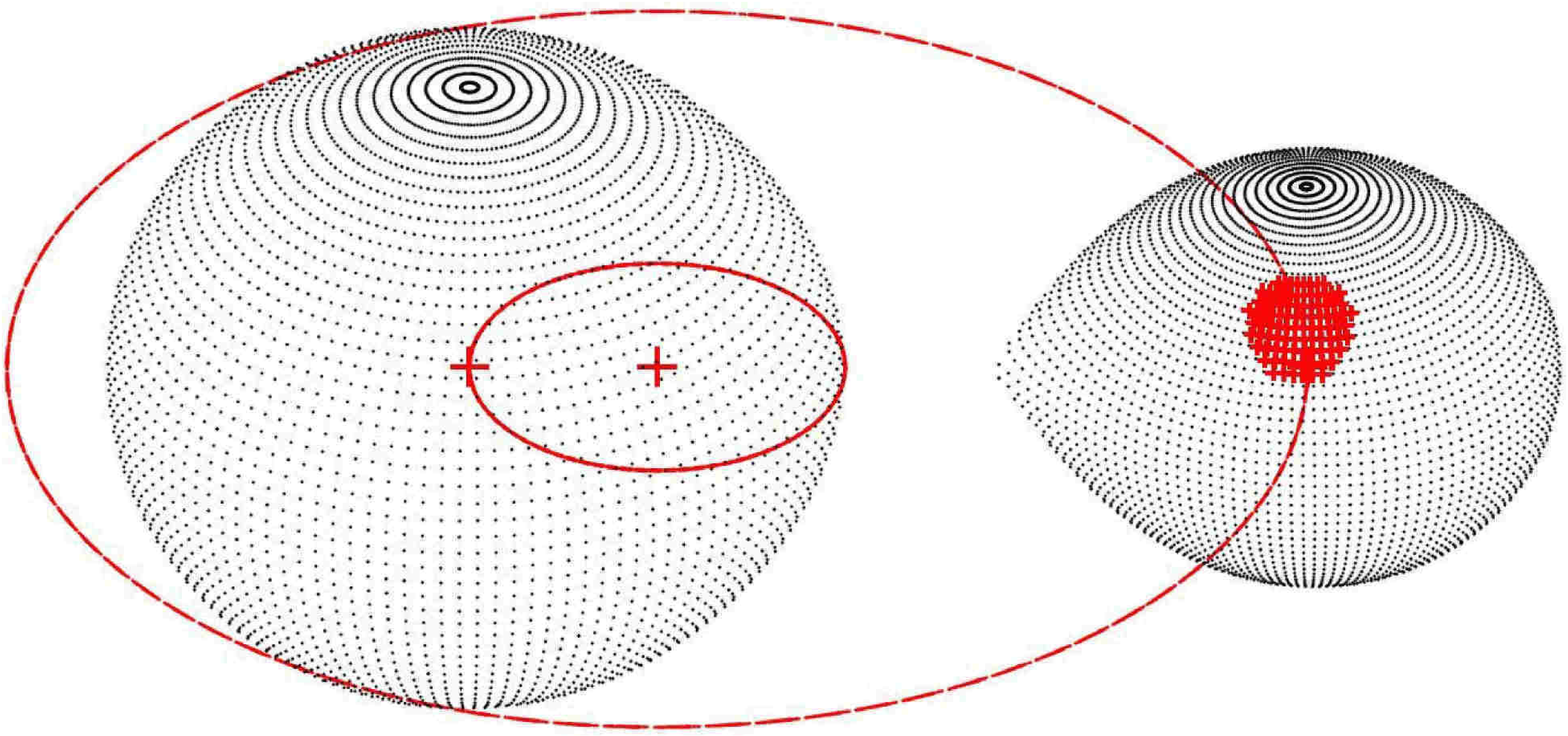}
\caption{phase 0.25}
\label{fig:subim1}
\end{subfigure}
\begin{subfigure}{0.5\textwidth}
\includegraphics[width=6.0cm, angle=0]{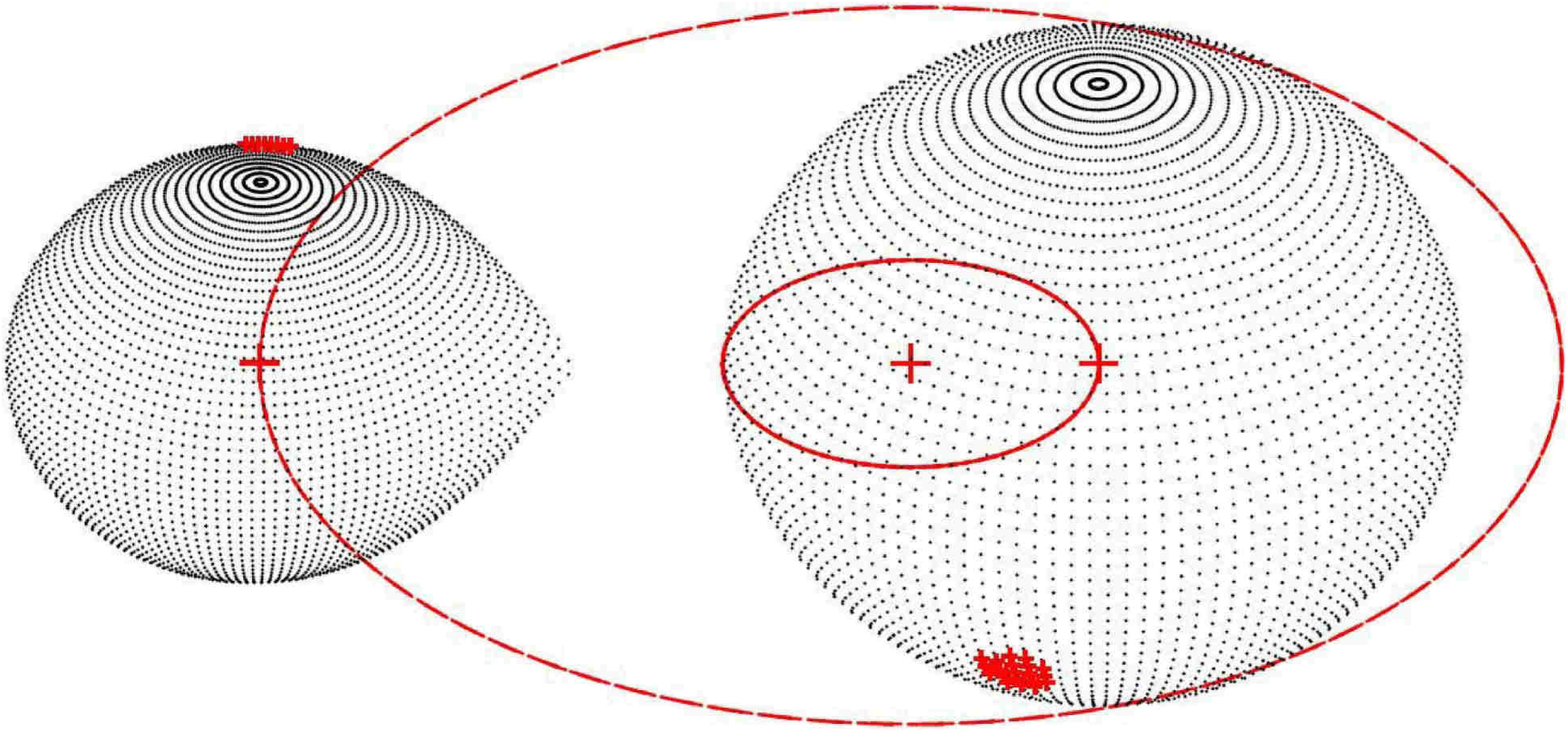}
\caption{phase 0.75}
\label{fig:subim2}
\end{subfigure}\\

\begin{subfigure}{0.5\textwidth}
\includegraphics[width=6.0cm, angle=0]{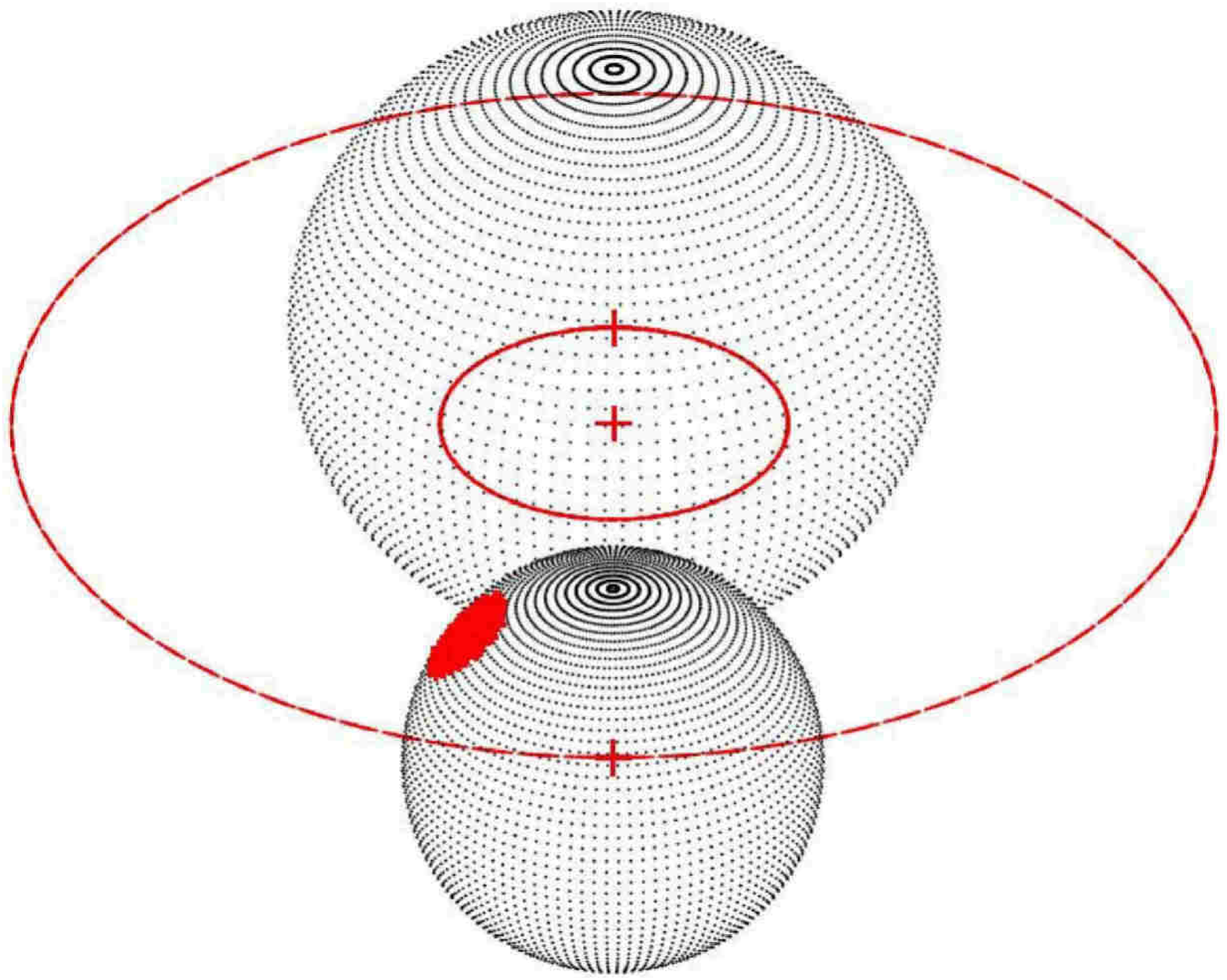}
\caption{phase 0.0}
\label{fig:subim3}
\end{subfigure}
\begin{subfigure}{0.5\textwidth}
\includegraphics[width=6.0cm, angle=0]{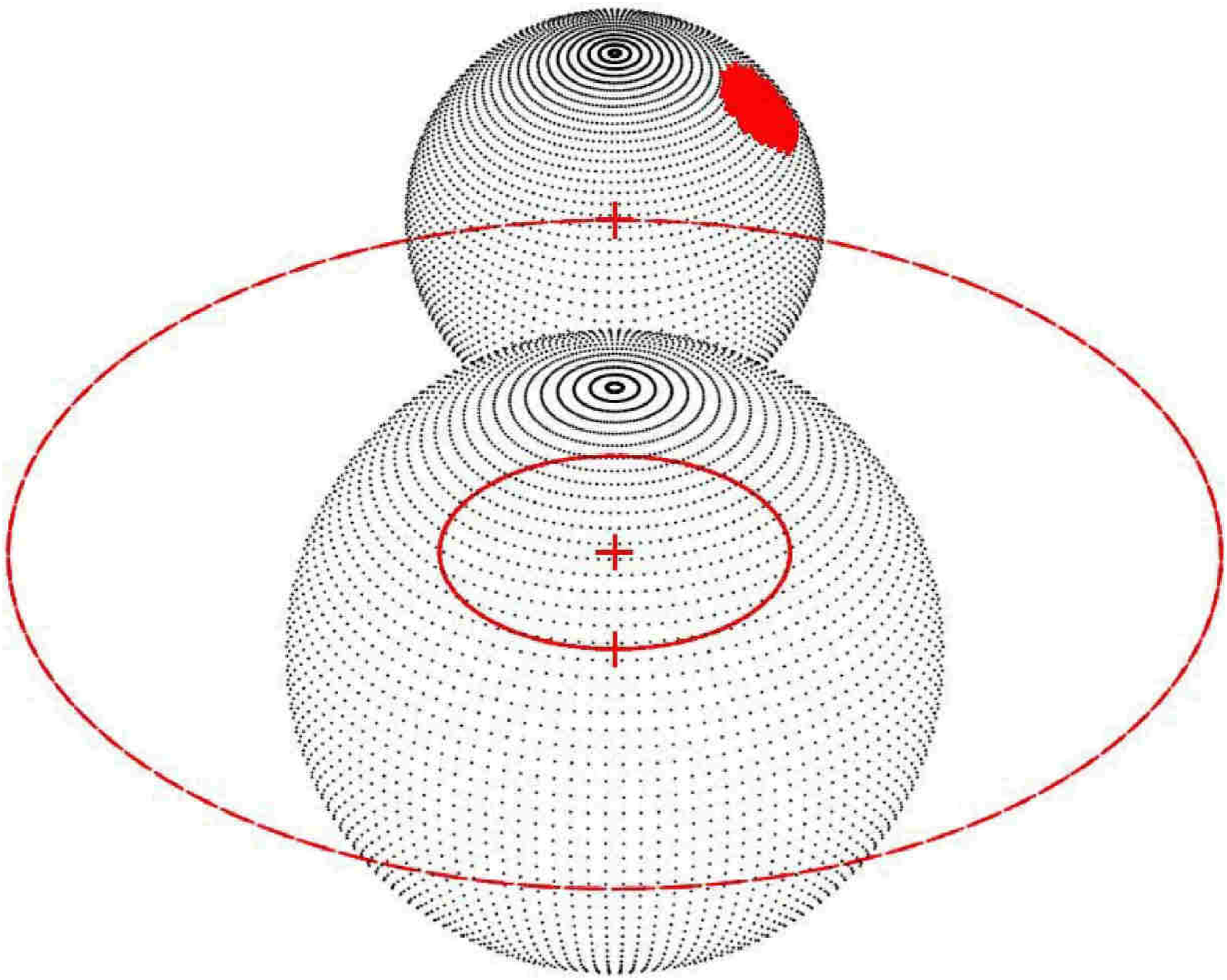}
\caption{phase 0.5}
\label{fig:subim4}
\end{subfigure}
\caption{The shape of the system KIC 2715417 at phases 0.0,~0.25,~0.5, and 0.75 \label{Fig9}. }
\end{figure}

\begin{figure}[h] 
  \centering
 \includegraphics[width=8.0cm, angle=0]{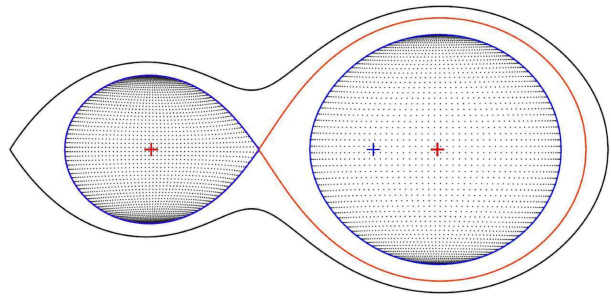}
 \begin{minipage}[]{85mm}
\caption{Roche Lobe Geometry of the system KIC 2715417\label{Fig10}. } 
\end{minipage}
\end{figure}

    Using the orbital and physical parameters listed in Table~\ref{tab2} with the Binary Maker 3 software \footnote{http://www.binarymaker.com/}, we present the shape of the system KIC 2715417 at phases 0.0, 0.25, 0.5, and 0.75 in Figure~\ref{Fig9}. Also we present the Roche lobe geometry of the system in Figure~\ref{Fig10}.

\begin{table}[h]
\bc
\begin{minipage}[]{100mm}
\caption[]{\small {The Orbital and Physical Parameters of the system KIC 2715417\label{tab2}}}
\end{minipage}
\setlength{\tabcolsep}{2.5pt}
\small
 \begin{tabular}{lccc}
  \hline\noalign{\smallskip}
Parameter~~~~&~~~~Value~~~~&~~~~Parameter~~~~&~~~~Value~~~~\\
  \hline\noalign{\smallskip}
Epoch & 2454964.666158 & $X_{1}$ & 0.655   \\
Period(day) & 0.2364399 &  $X_{2}$ & 0.711   \\
Inclination(i) & $56.56^{\circ} \pm 0.04$ &  $g_{1}$ & 0.32   \\
Mass ratio(q) & 0.289 ± 0.002 &  $g_{2}$ &0.32   \\
TAVH $(T_{1})$ & $5189^{\circ} K(Fixed)$ &  $A_{1}$ & 0.50   \\
TAVC $(T_{2})$ & $4711^{\circ} ± 14^{\circ} K$ &  $A_{2}$ & 0.50  \\
PHSV $(\Omega_{1})$ & $ 2.7635 \pm 0.0061 $ & $ \frac {L_{1}}{L_{1}+L_{2}} $ & $ 0.7909 \pm 0.0016 $\\
PHSV $(\Omega_{2})$ & 2.4416 $\pm$ 0.0032 & $\frac {L_{1}}{L_{1}+L_{2}}$ & 0.2091 $\pm$ 0.0030   \\
Phase Shift  & 0.00402 $\pm$ 0.00082 & $Residual \sum {(o-c)^{2}}$ & 0.0401   \\
Fill-out factor $(f_{1})$ & - 0.1150  &  $(f_{2}) $ & 0.0013 \\
$r_{1}(back)$ & 0.4716 $\pm$  0.0005 &  $r_{2}(back)$ &0.3109 $\pm$ 0.0024   \\
$r_{1}(side)$& 0.4625 $\pm$ 0.0004 & $ r_{2}(side)$ &0.2932 $\pm$ 0.0017   \\
$r_{1}(pole)$& 0.4505 $\pm$ 0.0003 & $ r_{2}(pole)$ &0.2764 $\pm$ 0.0015\\
$r_{1}(point)$ & 0.4859 $\pm$ 0.0008 &  $r_{2}(point)$ &0.3256 $\pm$ 0.0029   \\
Mean fractional radii ($r_{1}$) & 0.4616 $\pm$ 0.0004 & $r_{2}(Mean)$ & 0.2935 $\pm$ 0.0019  \\
\noalign{\smallskip}\hline
Spot Parameters & Spot 1 (primary) & Spot 2 (secondary )   \\
\noalign{\smallskip}\hline
Temp Factor & 0.86 $\pm$ 0.04 &  0.80 $\pm$ 0.05   \\
Spot Radius & $7^{\circ} \pm 2 $&  $14^{\circ} \pm 3$  \\
Longitude ($\lambda$) & $74^{\circ} \pm 12$ &  $87^{\circ} \pm 11$   \\
Co-Latitude ($\phi$) & $122^{\circ} \pm 8$ &  $46^{\circ} \pm 9$   \\
  \noalign{\smallskip}\hline
\end{tabular}
\ec
\end{table}
 \FloatBarrier
 

 \FloatBarrier
 
     Because of the O'Connell effect \citep{O'Connell+1951}, we applied the cool spot on the surface of star to achieve the best fit. We have one cool spot at each of the primary and secondary components for the system KIC 2715417 as shown in Figure~\ref{Fig9}. Table~\ref{tab2} gives the spot parameters where the spot of the secondary component has the largest radius and the coolest temperature than the primary component. 
     
    The temperature of the primary component is $5189^{\circ} K$ while the temperature factor of the cool spot is $0.86 \pm 0.04$. This means that the primary cool spot has a temperature of $4462^{\circ} \pm 208^{\circ} K$.On the other hand, the temperature of the secondary component is $4711^{\circ}+14^{\circ} K$ and the temperature factor of the cool spot is $0.80 \pm 0.05$. Thus the temperature of the cool spot on secondary component is $3769 ^{\circ} \pm 236^{\circ} K$.

\subsection{Light Curve Analyses of the System KIC 6050116 }

  The light curve of the system KIC 6050116 in the $V$-band as shown in Figure~\ref{Fig2} has been analyzed using the PHOEBE package. It is found that the corresponding temperature $T_{1}= 4569^{\circ} K$ from the Kepler Input Catalogue. With the PHOEBE software we have used the model of unconstrained binary system to estimate a set of parameters that represent the observed light curve. The best photometric fitting has been reached after several runs. 
  
       From the analyses of the software Binary Maker 3, the type of binary system is considered to be overcontact, since the fill-out factors of the first component is $f_{1}$= 0.0704 and the secondary component is $f_{2}$= 0.0704. The fill-out factor is equal for the two components of contact and overcontact binaries as the two stars are in contact or overcontact with each other. They must have the same gravitational equipotential $(\Omega)$ otherwise gas will literally leak away from the system until it reaches equilibrium.
       
\begin{figure}[h] 
  \centering
 \includegraphics[width=8.0cm, angle=0]{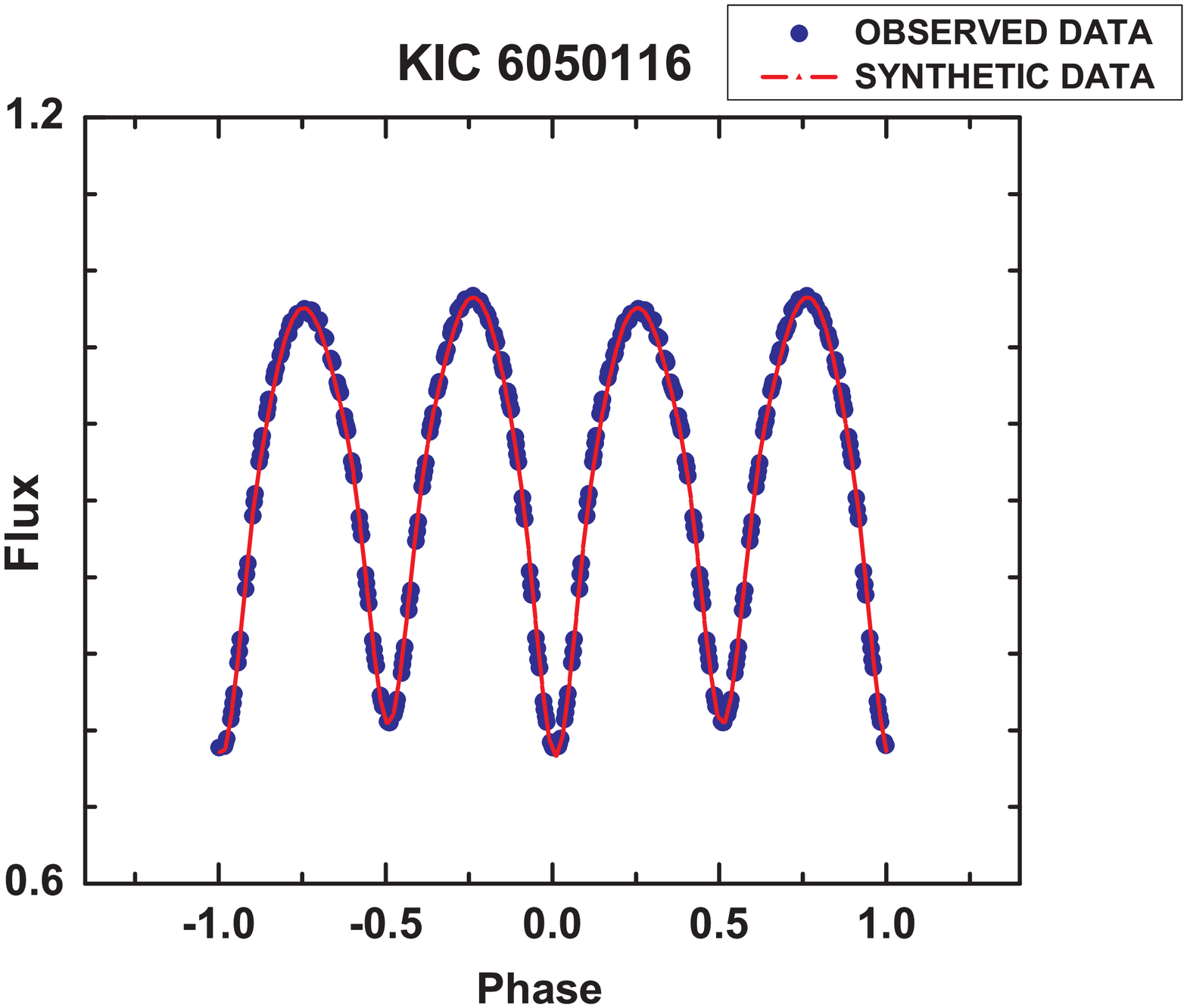}
 \begin{minipage}[]{85mm}
\caption{Light curve in $V$ filter showing the coherence between observed and synthetic curves\label{Fig11}. } 
\end{minipage}
\end{figure}

\begin{figure}[h] 
  \centering
 \includegraphics[width=8.0cm, angle=0]{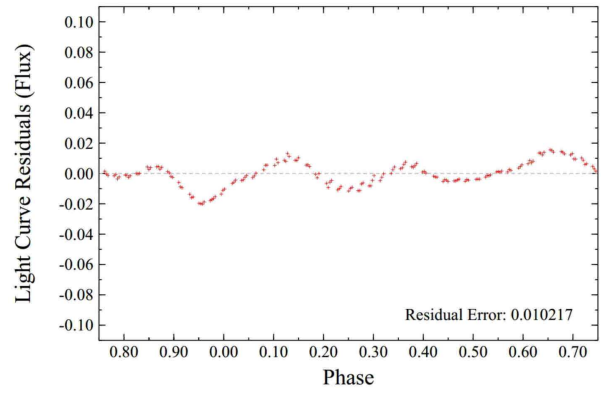}
 \begin{minipage}[]{85mm}
\caption{Light curve residual error for different phases for the system KIC 6050116\label{Fig12}. } 
\end{minipage}
\end{figure}

      The primary component is massive and hotter than the secondary one, with a temperature difference of about $84^{\circ}$ K.The orbital and physical parameters of the system KIC 6050116 are listed in Table~\ref{tab3}. Figure~\ref{Fig11} displays the observed light curve for the interval 2454964.51229 JD - 2454974.48433 JD, together with the synthetic curve in the V band while Figure~\ref{Fig12} displays the light curve residual error for different phases. 
      
\begin{figure}[h]
\begin{subfigure}{0.5\textwidth}
\includegraphics[width=6.0cm, angle=0]{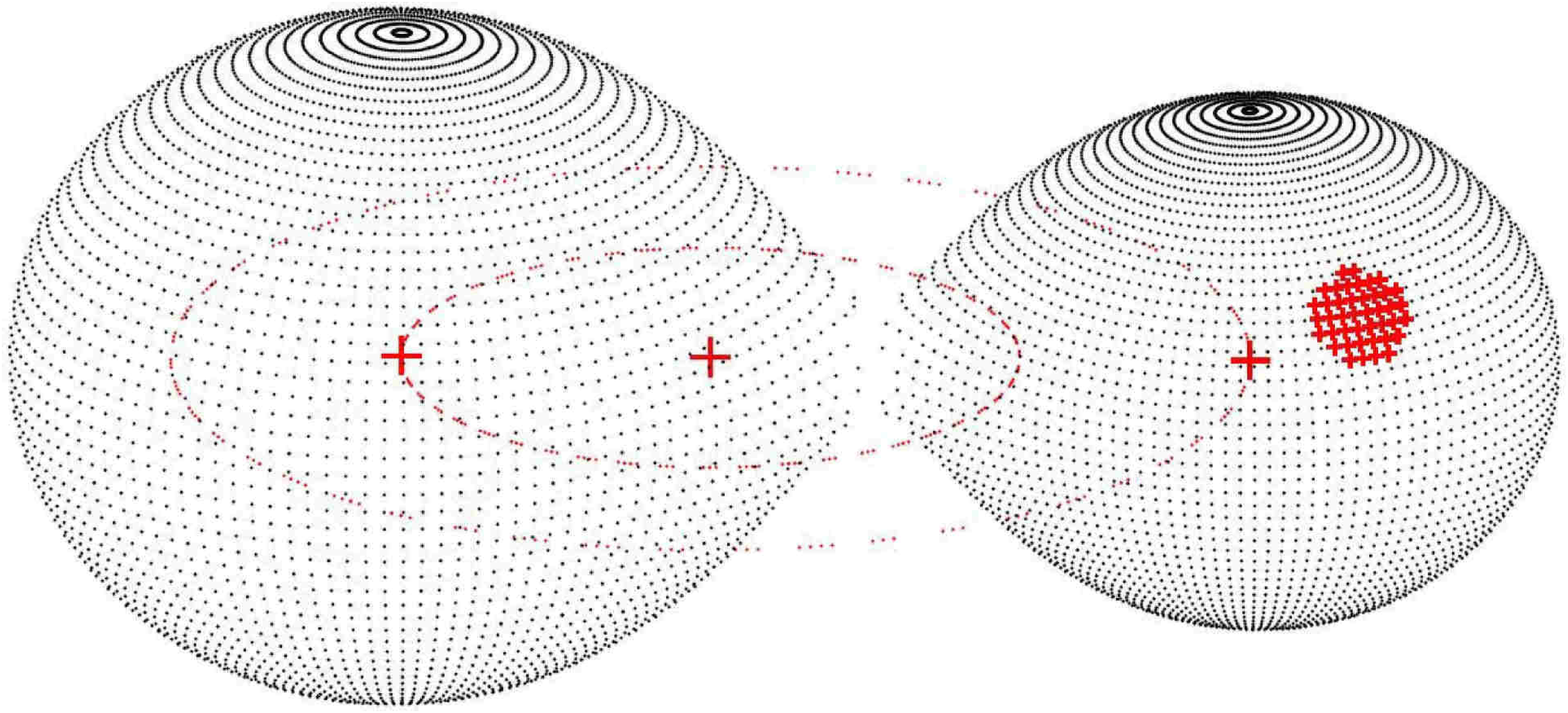}
\caption{phase 0.25}
\label{fig:subim5}
\end{subfigure}
\begin{subfigure}{0.5\textwidth}
\includegraphics[width=6.0cm, angle=0]{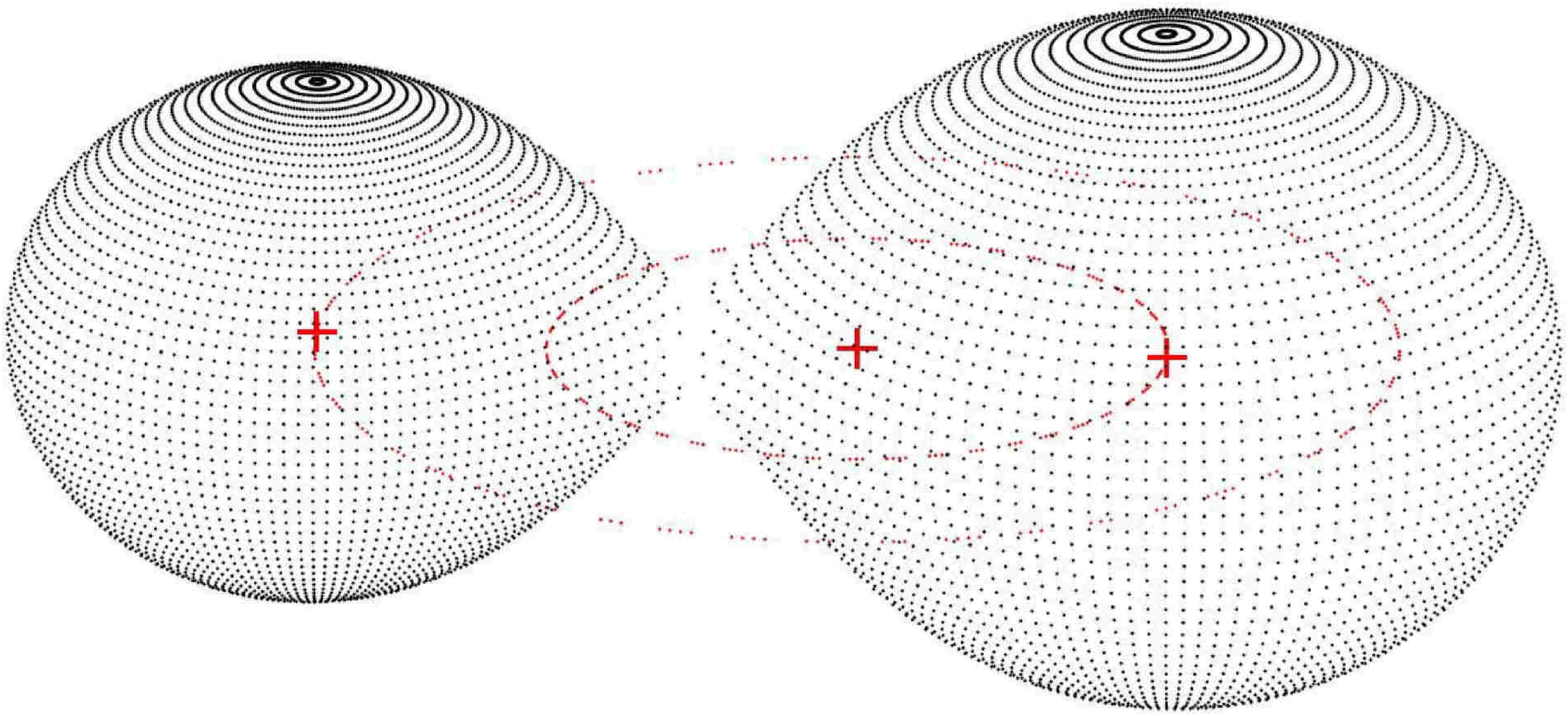}
\caption{phase 0.75}
\label{fig:subim6}
\end{subfigure}\\

\begin{subfigure}{0.5\textwidth}
\includegraphics[width=6.0cm, angle=0]{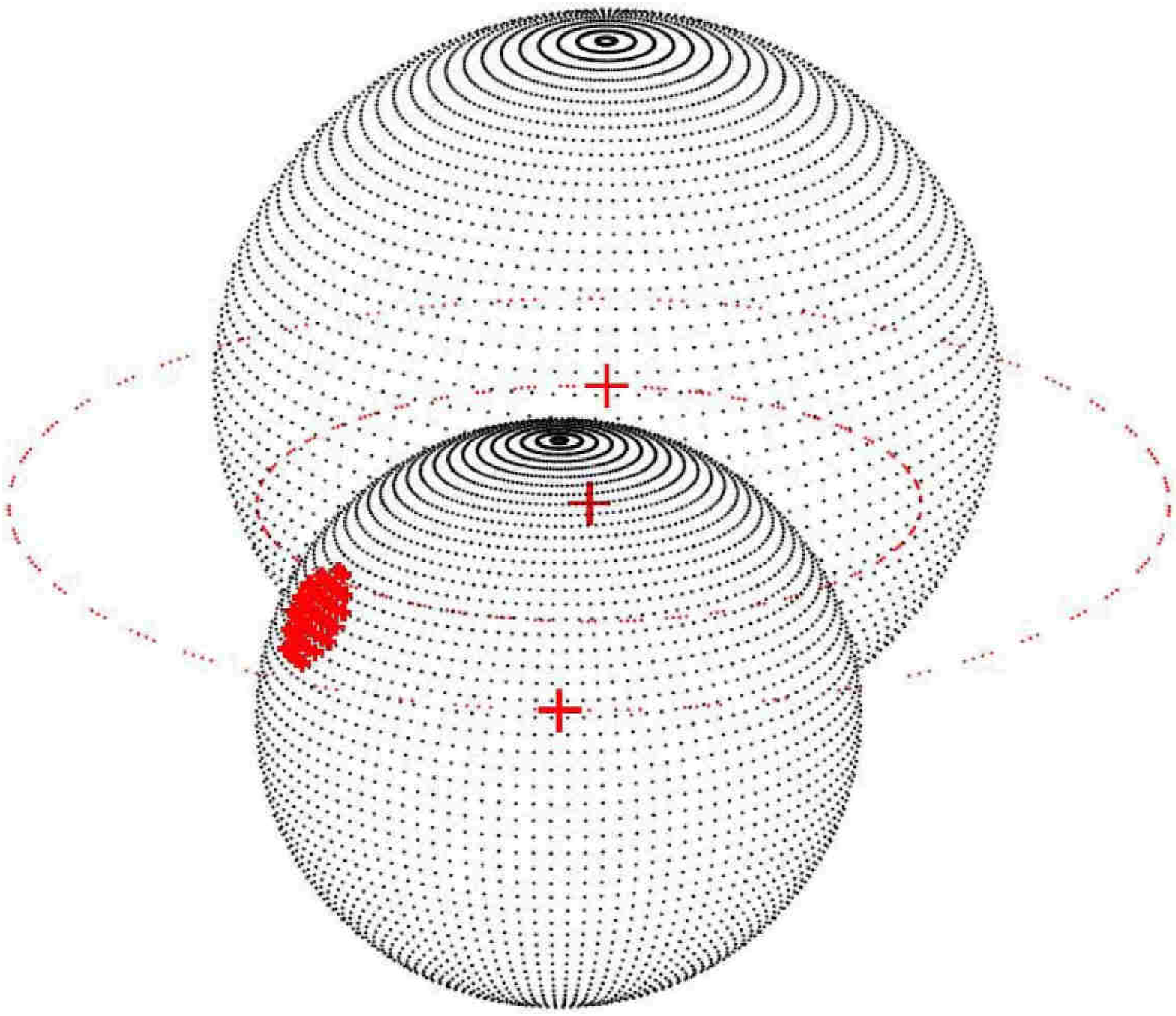}
\caption{phase 0.0}
\label{fig:subim7}
\end{subfigure}
\begin{subfigure}{0.5\textwidth}
\includegraphics[width=6.0cm, angle=0]{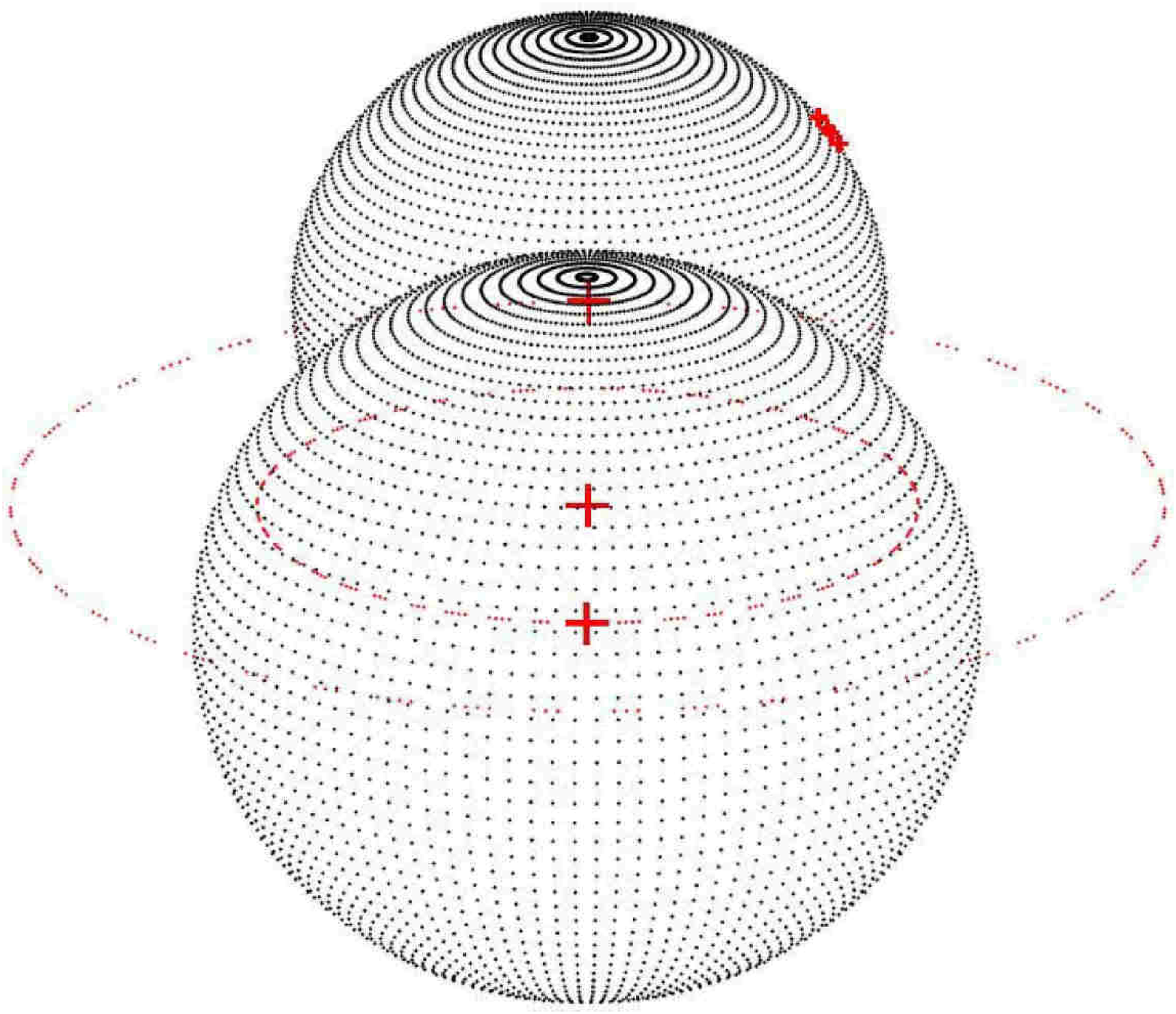}
\caption{phase 0.5}
\label{fig:subim8}
\end{subfigure}
\caption{The shape of the system KIC 6050116 at phases 0.0,~0.25,~0.5, and 0.75 \label{Fig13}. }
\end{figure}

\begin{figure}[h] 
  \centering
 \includegraphics[width=8.0cm, angle=0]{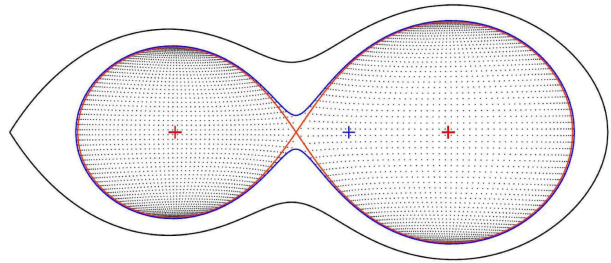}
 \begin{minipage}[]{85mm}
\caption{Roche Lobe Geometry of the system KIC 6050116\label{Fig14}. } 
\end{minipage}
\end{figure}

     According to the effective temperature of both the primary and secondary components of the system KIC 6050116 and from the calibration of Morgan-Keenan (MK) spectral types for the main sequence class, the spectral types are nearest to $K_{5}$ for both components. Using the orbital and physical parameters listed in Table~\ref{tab3} with the BM3 program, we present the shape of the system KIC 6050116 at phases 0.0, 0.25, 0.5, and 0.75 in Figure~\ref{Fig13}. We also present the Roche lobe geometry of the system in Figure~\ref{Fig14}.

\begin{table}[h]
\bc
\begin{minipage}[]{100mm}
\caption[]{\small {The Orbital and Physical Parameters of the system KIC 6050116\label{tab3}}}
\end{minipage}
\setlength{\tabcolsep}{2.5pt}
\small
 \begin{tabular}{lccc}
  \hline\noalign{\smallskip}
Parameter~~~~&~~~~Value~~~~& ~~~~Parameter~~~~&~~~~Value~~~~\\
  \hline\noalign{\smallskip}
Epoch & 2454964.708006 & $X_{1}$ & 0.787   \\
Period(day) & 0.2399081 &  $X_{2}$ & 0.795   \\
Inclination(i) & $69.2^{\circ} \pm 0.7$ &  $g_{1}$ & 0.32   \\
Mass ratio(q) & 0.573 ± 0.012 &  $g_{2}$ &0.32   \\
TAVH $(T_{1})$ & $4569^{\circ} K(Fixed)$ &  $A_{1}$ & 0.50   \\
TAVC $(T_{2})$ & $4485^{\circ} ± 14^{\circ} K$ &  $A_{2}$ & 0.50  \\
PHSV $(\Omega_{1})$ & $ 2.9899 \pm 0.0016 $ & $ \frac {L_{1}}{L_{1}+L_{2}} $ & $ 0.6504 \pm 0.0020 $\\
PHSV $(\Omega_{2})$ & 2.9899 $\pm$ 0.0016 & $\frac {L_{1}}{L_{1}+L_{2}}$ & 0.3496 $\pm$ 0.0037   \\
Phase Shift  & 0.00857 $\pm$ 0.00035 & $ \sum {(o-c)^{2}}$ & 0.107112   \\
$(f_{1})$ & 0.0704  &  $(f_{2}) $ & 0.0704 \\
$r_{1}(back)$ & 0.4618 $\pm$ 0.0040 &  $r_{2}(back)$ &0.3637 $\pm$ 0.0084   \\
$r_{1}(side)$ & 0.4315 $\pm$  0.0025 &  $r_{2}(side)$ &0.3291 $\pm$ 0.0062  \\
$r_{1}(pole)$& 0.4067 $\pm$ 0.0018 & $ r_{2}(pole)$ &0.3144 $\pm$ 0.0053   \\
$r_{1}(point)$& 0.5570 $\pm$ 0.0021 & $ r_{2}(point)$ &0.4430 $\pm$ 0.0021\\
$r_{1}(Mean)$ & 0.4245 $\pm$ 0.0028 & $r_{2}(Mean)$ & 0.3337 $\pm$ 00045  \\
\noalign{\smallskip}\hline
Spot Parameters &  Spot (secondary )   \\
\noalign{\smallskip}\hline
Temp Factor & 0.76 $\pm$ 0.04  \\
Spot Radius & $10^{\circ} \pm 3 $  \\
Longitude ($\lambda$) & $62^{\circ} \pm 7$    \\
Co-Latitude ($\phi$) & $117^{\circ} \pm 9$   \\
  \noalign{\smallskip}\hline
\end{tabular}
\ec
\end{table}
 \FloatBarrier


 \FloatBarrier
 
We have also one cool spot at secondary component for the system KIC 6050116 as shown in Figure~\ref{Fig13}. Table~\ref{tab3} gives the temperature of the secondary component as $4485^{\circ} \pm 17^{\circ}$ K and the temperature factor of the cool spot is $0.76~\pm~0.04$. Thus the temperature of the cool spot of secondary component is $3409^{\circ} \pm 179^{\circ}$ K.

\subsection{Light Curve Analyses of the System KIC 6287172}

The light curve of the system KIC 6287172 in the $V$-band as shown in Figure~\ref{Fig3} has been analyzed using the PHOEBE package. We found that the corresponding temperature $T_{1}= 6646^{\circ}$ K from the Kepler Input Catalogue. With the PHOEBE software we have used the model of unconstrained binary to determine approximate values of the physical and geometrical parameters that represent the observed light curve. After the best photometric fitting has been reached, we have used the data obtained from PHOEBE software inserted in the other software Binary Maker 3 to find the type of binary system from the fill-out factors of the two components. 

\begin{figure}[h] 
  \centering
 \includegraphics[width=8.0cm, angle=0]{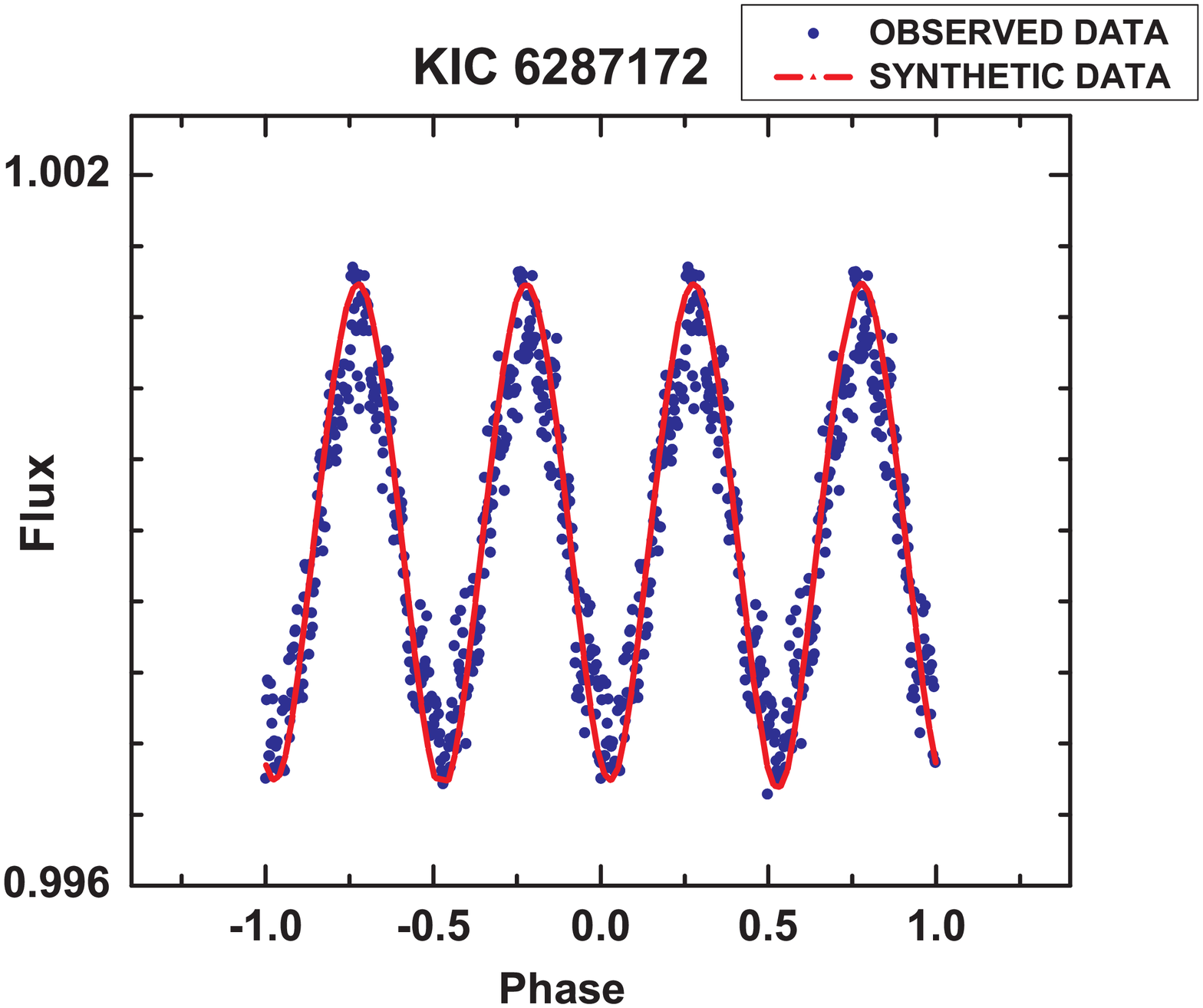}
 \begin{minipage}[]{85mm}
\caption{Light curve in $V$ filter showing the coherence between observed and synthetic curves\label{Fig15}. } 
\end{minipage}
\end{figure}

\begin{figure}[h] 
  \centering
 \includegraphics[width=8.0cm, angle=0]{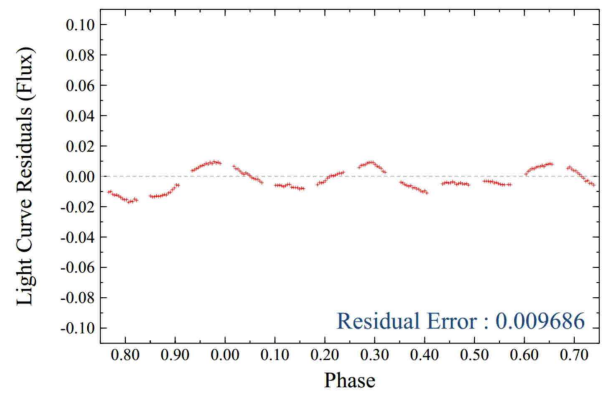}
 \begin{minipage}[]{85mm}
\caption{Light curve residual error for different phases\label{Fig16}. } 
\end{minipage}
\end{figure}

   We found that the fill-out parameters of the primary star is $f_{1}$= 0.432657 and for the secondary star is $f_{2}$= 0.432657 which is considered a non-eclipsing overcontact type or Ellipsoidal Variables $(ELV)$.  The primary component is massive and hotter than the secondary one, with a temperature difference of about $22^{\circ}$ K. The orbital and physical parameters of the system KIC 6287172 are listed in Table~\ref{tab4}. Figure~\ref{Fig15} displays the observed light curve for the interval 2454953.53905 JD - 2454967.31191 JD, together with the synthetic curve in the $V$-band, while Figure~\ref{Fig16} represents the light curve residual error for different phases.
   
\begin{figure}[h]
\begin{subfigure}{0.5\textwidth}
\includegraphics[width=6.0cm, angle=0]{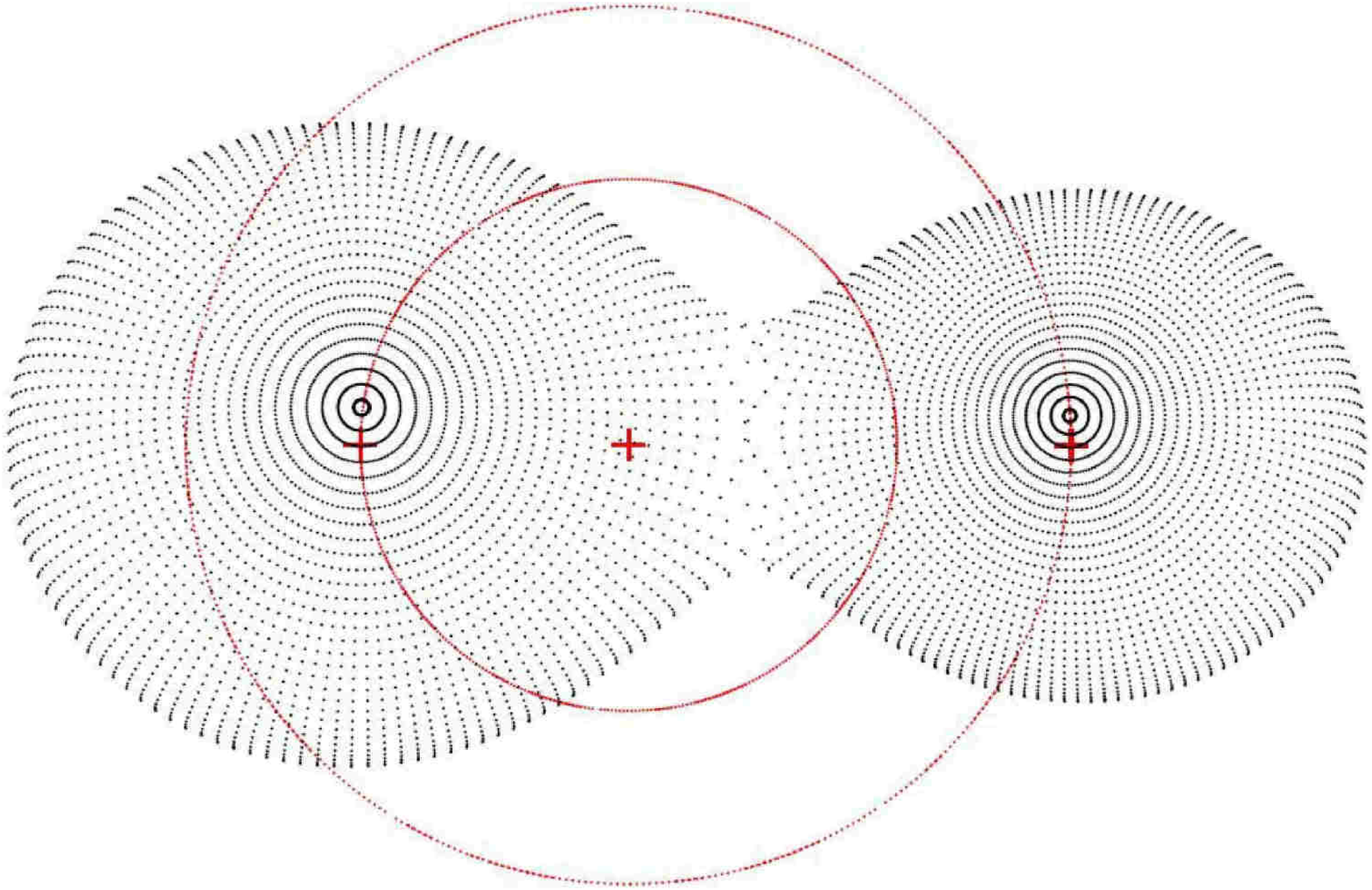}
\caption{phase 0.25}
\label{fig:subim9}
\end{subfigure}
\begin{subfigure}{0.5\textwidth}
\includegraphics[width=6.0cm, angle=0]{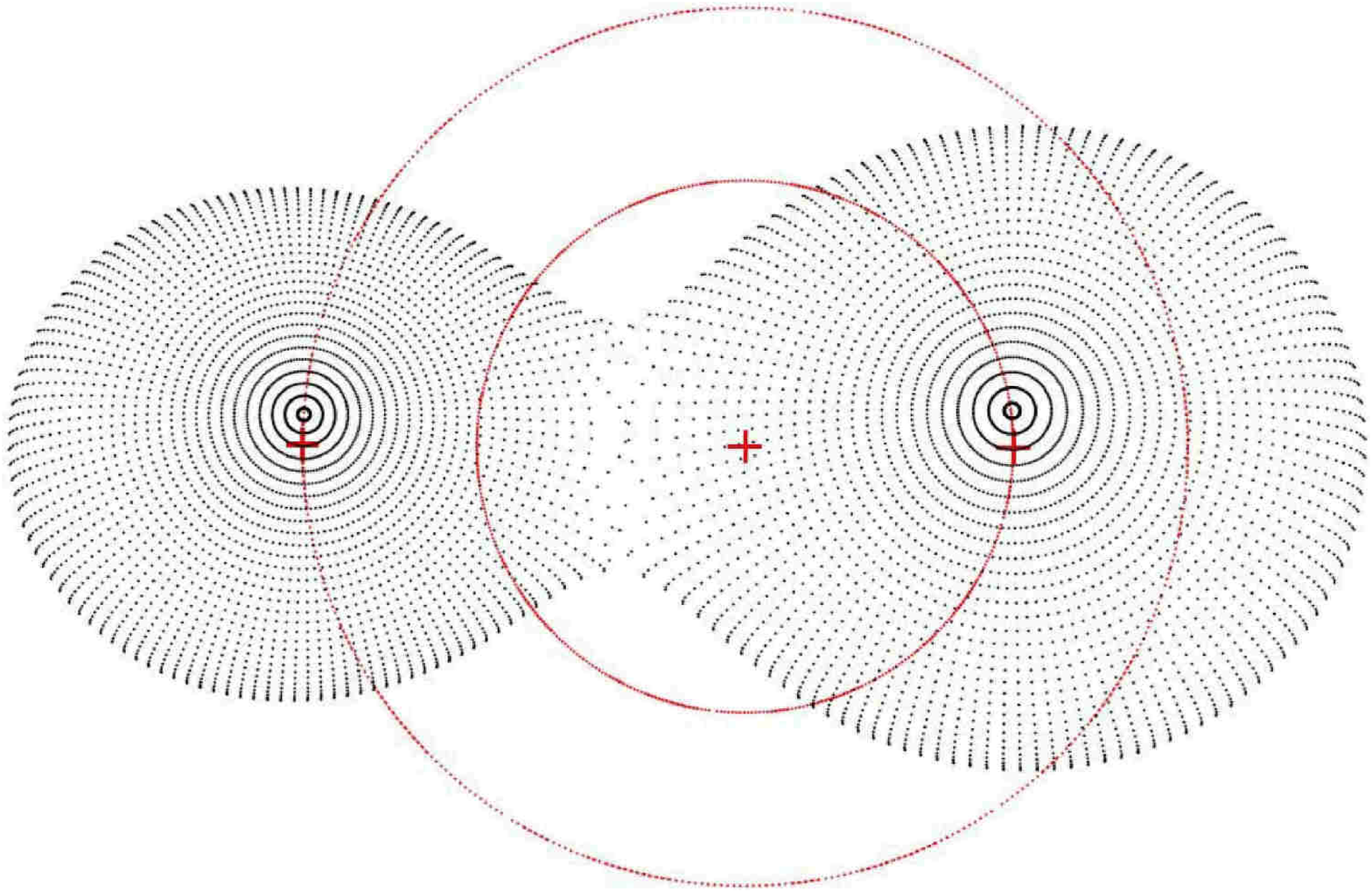}
\caption{phase 0.75}
\label{fig:subim10}
\end{subfigure}\\

\begin{subfigure}{0.5\textwidth}
\includegraphics[width=6.0cm, angle=0]{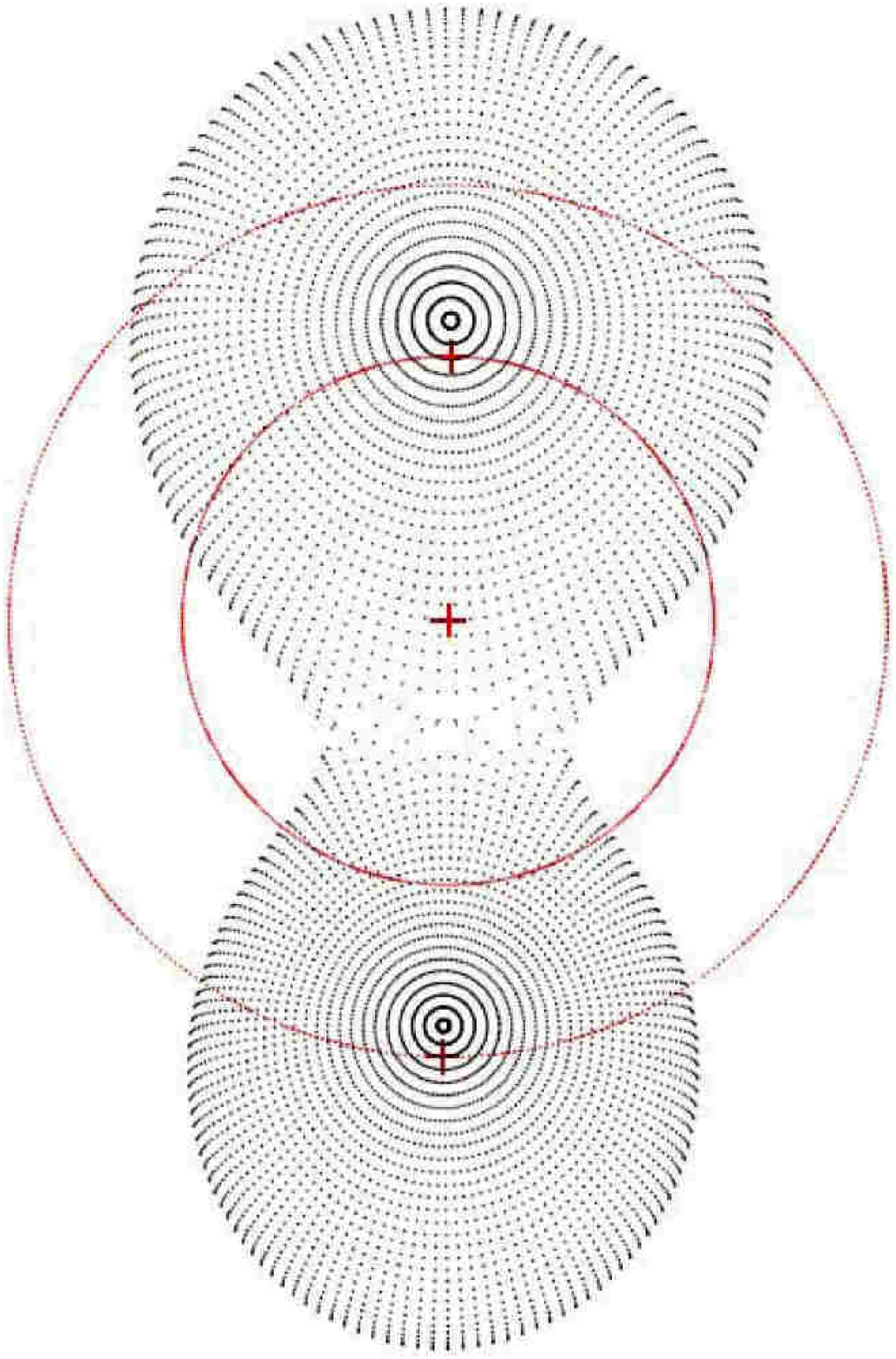}
\caption{phase 0.0}
\label{fig:subim11}
\end{subfigure}
\begin{subfigure}{0.5\textwidth}
\includegraphics[width=6.0cm, angle=0]{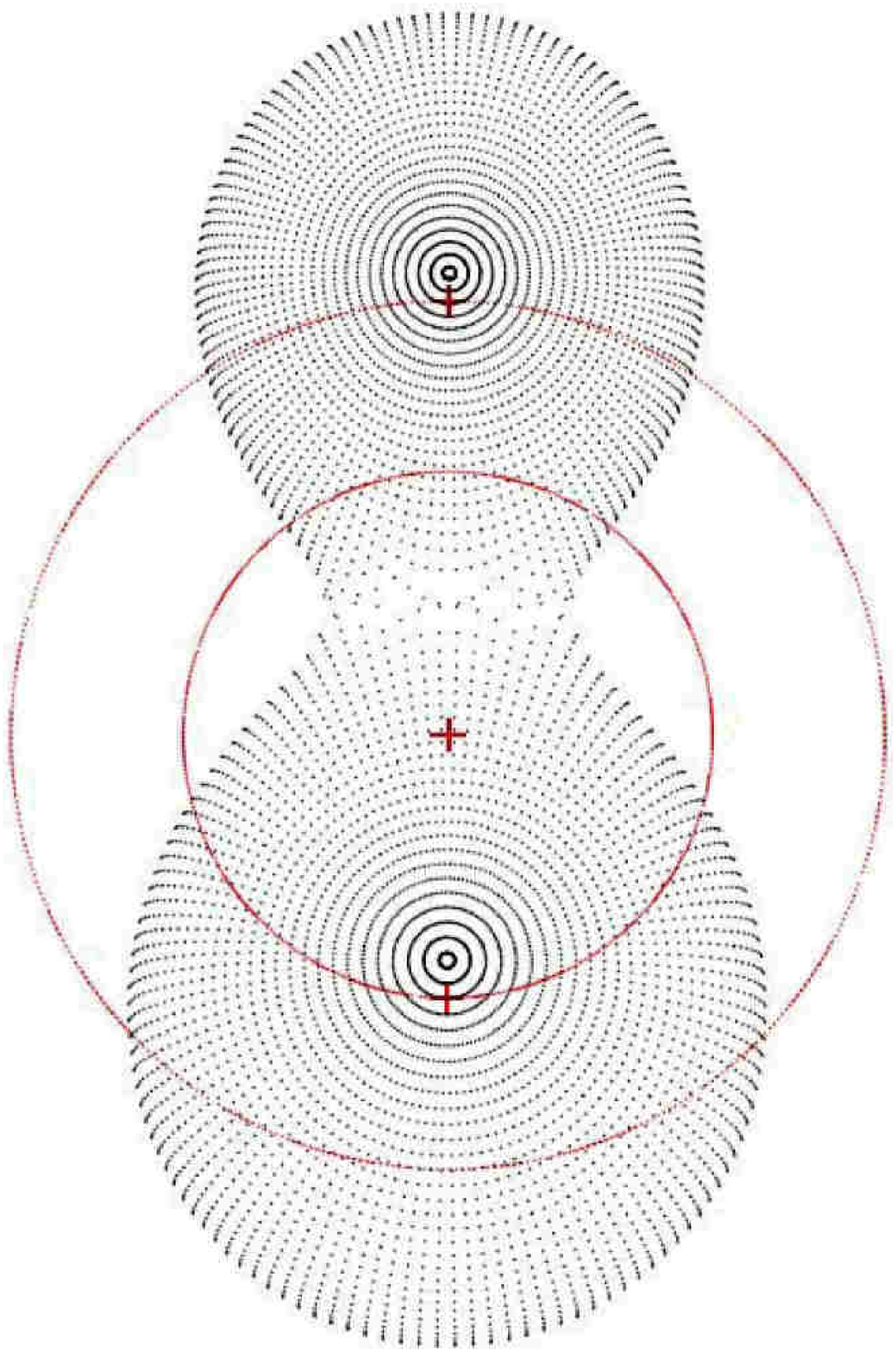}
\caption{phase 0.5}
\label{fig:subim12}
\end{subfigure}
\caption{The shape of the system KIC 6287172 at phases 0.0,~0.25,~0.5, and 0.75 \label{Fig17}. }
\end{figure}

\begin{figure}[h]
  \centering
 \includegraphics[width=8.0cm, angle=0]{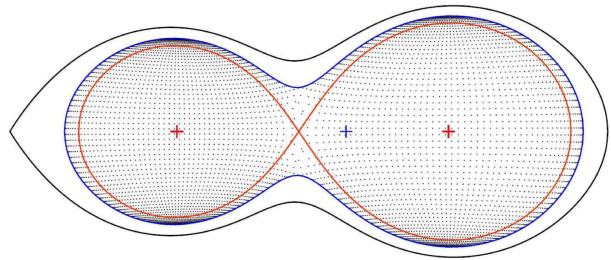}
 \begin{minipage}[]{85mm}
\caption{Roche Lobe Geometry of the system KIC 6287172\label{Fig18}. } 
\end{minipage}
\end{figure}

    According to the effective temperature of both the primary and secondary components of the system KIC 6287172 and from the calibration of Morgan-Keenan (MK) spectral types for the main sequence class, the spectral types are nearest to $F_{5}$ for both components. Using the orbital and physical parameters listed in Table~\ref{tab4} with the BM3 program, we present the shape of the system KIC 6287172 at phases 0.0, 0.25, 0.5, and 0.75 in Figure~\ref{Fig17}. We present the Roche geometry of the system in Figure~\ref{Fig18}. 

\begin{table}[h]
\bc
\begin{minipage}[]{100mm}
\caption[]{\small {The Orbital and Physical Parameters of the system KIC 6287172\label{tab4}}}\end{minipage}
\setlength{\tabcolsep}{2.5pt}
\small
 \begin{tabular}{lccc}
  \hline\noalign{\smallskip}
Parameter~~~~&~~~~Value~~~~&~~~~Parameter~~~~&~~~~Value~~~~\\
  \hline\noalign{\smallskip}
Epoch & 2454953.651911 & $X_{1}$ & 0.507   \\
Period(day) & 0.2038732 &  $X_{2}$ & 0.509  \\
Inclination(i) & $7.1^{\circ} \pm 0.2$ &  $g_{1}$ & 0.32   \\
Mass ratio(q) & 0.606 ± 0.001 &  $g_{2}$ &0.32   \\
TAVH $(T_{1})$ & $6646^{\circ} K(Fixed)$ &  $A_{1}$ & 0.50   \\
TAVC $(T_{2})$ & $6624^{\circ} ± 11^{\circ} K$ &  $A_{2}$ & 0.50  \\
PHSV $(\Omega_{1})$ & $ 2.9213 \pm 0.0037 $ & $ \frac {L_{1}}{L_{1}+L_{2}} $ & $ 0.6137 \pm 0.0062 $\\
PHSV $(\Omega_{2})$ & 2.9213 $\pm$ 0.0037 & $\frac {L_{1}}{L_{1}+L_{2}}$ & 0.3863  $\pm$ 0.0062  \\
Phase Shift  & 0.02983 $\pm$ 0.00031 & $ \sum {(o-c)^{2}}$ & 0.475962   \\
$(f_{1})$ & 0.4327  &  $(f_{2}) $ & 0.4327 \\
$r_{1}(back)$ & 0.4954  $\pm$ 0.0004 &  $r_{2}(back)$ &0.4133 $\pm$ 0.0008   \\
$r_{1}(side)$  & 0.4537 $\pm$  0.0002 &  $r_{2}(side)$ &0.3609 $\pm$ 0.0005  \\
$r_{1}(pole)$& 0.4232 $\pm$ 0.0002 & $ r_{2}(pole)$ &0.3409 $\pm$ 0.0004  \\
$r_{1}(point)$& 0.5513 $\pm$ 0.0002 & $ r_{2}(point)$ &0.4487 $\pm$ 0.0002\\
$r_{1}(Mean)$ & 0.4245 $\pm$ 0.0028 & $r_{2}(Mean)$ & 0.3337 $\pm$ 00045  \\
\noalign{\smallskip}\hline
\end{tabular}
\ec
\end{table}
 \FloatBarrier


 \FloatBarrier

\section{ABSOLUTE PARAMETERS OF THE SYSTEMS}

The physical parameters: effective temperature $T_{eff}$, absolute magnitude $M_{V}$, relative radius $\frac {R}{R_{\odot}}$, relative luminosity $\frac {L}{L_{\odot}}$, and surface gravity g, were calculated using the equations of stellar structure. The adopted constants which we used are $T_{eff \odot}=5772^{\circ} K$, $\log g_{\odot}=4.44$, and $M_{bol\odot}= 4.74$.

We have three methods to determine only the primary masses for each type of binary. These three methods are: \cite{Harmanec+1998}, \cite{Maceroni+Van'tVeer+1996}  and \cite{Gazeas+Niarchos+2006}.
 \subsection {Harmanec’s method:}
 
  We compute the physical parameters according to the empirical relation derived by \cite {Harmanec+1998} from his work about the stellar masses and radii based on modern binary data. These relation obtained by the least-square fit via Chebychev polynomials for the data introduced by \cite {Popper+1980}. Harmanec’s method is the mass - temperature relation used to determine the masses of the primary and secondary components $M_{1}$ and $M_{2}$ for detached binaries. However, in the case of overcontact binaries and any other type of binaries, it is only applicable for determination of the primary component $M_{1}$ which is slightly affected by thermal contact with the secondary low mass stars. If we assumed that the effect of temperature of the secondary component on the primary is negligible because the primary has higher temperature. Thus the primary can be deduced from the previous relations with good approximation. The mass of the secondary component can’t accurately be calculated using this method because the secondary component is greatly affected by thermal contact with the primary hotter star. 
The mass of primary stars according to Harmanec were calculated from the following equation:
\begin{equation}
\log {\frac {M_{1}}{M_{\odot}}}=((1.771141{X}-21.46965){X}+88.05700){X}-121.6782 
 \label{eq4}
\end{equation}

Where  ~~~~~~~~~~~~~~~~~~~~~~~~~~~~~~~~~~~~~~~~~~ $X = \log (T_{eff1})$\\
Equation \ref{eq4} is applicable for   $ 4.62 \ge  \log (T_{eff})  \ge 3.71 $\\
The mass of secondary component is determined from the photometric mass ratio, $(q=\frac {M_{2}}{M_{1}})$
 \subsection {Maceroni \& Van't Veer's method:}

The physical parameters can be determined from the relation between the total mass $(M_{T})$ and total luminosity $(L_{T})$ of the binary systems from \cite{Maceroni+Van'tVeer+1996} taking the assumption that the interaction between the two components of the binary system is not affecting the total luminosity of the system, therefore the common envelope radiates the luminosity given by the sum of the internal luminosities In other words, the total luminosity is the same for the binary system. Due to the pervious assumption, this method can be applicable for any type of binary system such as detached, over contact or ellipsoidal to determine the total mass of the binary system. Using the value of mass ratio for each binary, we can get the individual masses $M_{1}$ and $M_{2}$.for the binary system as shown in the following equations:

\begin{equation}
\log (L_{T})=\frac {2}{3} \log (M_{T})+c_{1} 
\label{eq5}
\end{equation}  
~~~~~~~~Where ~~~~~~~~~~~~~~~~~~~~~~~~~~~~~~~~~~~~~~~~~~~$c_{1}=\log \left [ c{P^{(\frac{4}{3})}}({r_{1}^{2}}{T_{1}^{4}}+{r_{2}^{2}}{T_{2}^{4}})\right ]$\\
In the same paper \citep{Maceroni+Van'tVeer+1996} at equation (2), the constant c was written as:

~~~~~~~~~~~~~~~~~~~~~~~~~~~~~~~~~~~~~~~~~~~~~~~~~~~~~~~~~$ c={({4 \pi})^{\frac{1}{3}}}~~{G^{\frac{2}{3}}}~~{\sigma}$\\
But we note that, there is a mistake in the formula of c and it should be written as:

~~~~~~~~~~~~~~~~~~~~~~~~~~~~~~~~~~~~~~~~~~~~~~~~~~~~~~~~~$ c={(\frac{4}{\pi})^{\frac{1}{3}}}~~{G^{\frac{2}{3}}}~~{\sigma}$\\
Using the evolutionary tracks for non-rotating model which have been computed by \cite {Mowlavi+etal+2012} for zero age main sequence stars (ZAMS) with metallicity $Z=0.014$. We correlate the total luminosity and total mass from the fitting data of the binary system of stars located on the ZAMS as follows:

\begin{equation}
\log (L_{T})=4.3 \log (M_{T}) - 0.539
 \label{eq6}
\end{equation}  

From the intersections between the straight lines of binary systems represented by equations (\ref{eq5}) and (\ref{eq6}) as shown in Figures \ref{Fig19}, \ref{Fig23} and \ref{Fig27}, we can deduced the total mass $(M_{T})$ of binary system. From the knowledge of the mass ratio $(q)$ and total mass $(M_{T})$ of the binary system, the individual masses $(M_{1}~~and~~M_{2})$ can readily be calculated from equation \ref{eq7}. 

\begin{equation}
where~~~~~~~~~~~~~~~~~~~~~~~~~~~~~M_{1}=\frac {M_{T}}{q+1},~~~~~~~~~~~M_{2} = M_{1} \times q =\frac {{M_{T}}\times q}{(q+1)}~~~~~~~~~~~~~~~~~~~~~~~
\label{eq7}
\end{equation}  
 \subsection {Gazeas \& Niarchos' method:}

This is the period - mass relation used to determine the mass of the primary component.  It is only applicable for short period eclipsing binary systems such as overcontact or near contact types with orbital period $ \log P < - 0.25$ and is not included Ellipsoidal variables (non-eclipsing binaries) according to \cite{Gazeas+Niarchos+2006} and \cite{Gazeas+Stepien+2008}.  The mass of the primary star $(M_{1})$ can be obtained from the following expression:
\begin{equation}
\log (M_{1})=(0.755 \pm 0.059)\log (P) + (0.416 \pm 0.024) 
\label{eq8}
\end{equation}

Where P is the orbital period of the binary system of W UMa-type. 
All of these three methods are used to calculate only the mass of the primary component of the binary system. The mass of the secondary component of binary system is found from the photometric mass ratio $(q =\frac {M_{2}}{M_{1}})$ and the primary mass. 
The photometric mass ratio of the contact type is more precise than undercontact type such as the detached binary. All of these three methods are used to calculate only the individual masses of the binary systems. From the Kepler third law, the semi-major axis of the orbit for binary system was determined using the following equation:

\begin{equation}
a=\sqrt[3]{\frac{G\times{(M_{1}+M_{2})}\times{P^{2}}}{4{\pi^{2}}}}
\label{eq9}
\end{equation}

Where $G$ is the gravitational constant $(G= 6.67428 \times 10^{-11} kg^{-1} m^{3} s^{-2})$.
Knowing the semi-major axis $(a)$ thus, we can calculate the radius of primary and secondary stars according to the relation  $R_{j}=r_{j}×a$ , where $j$ can take 1 for primary component or 2 for secondary component, $r$ is the mean fractional radii of the star.
The luminosity of the primary and secondary stars were calculated using the direct equation:

\begin{equation}
\frac {L_{j}}{L_{\odot}}=(\frac {R_{j}}{R_{\odot}})^{2} \times (\frac {T_{j}}{T_{\odot}})^{4}
\label{eq10}
\end{equation}

The bolometric magnitude of the primary and secondary components of binary systems is calculated using the following equation:

\begin{equation}
 {M_{bol j}}={M_{bol \odot}} - 2.5 \log_{10} \frac {L_{j}}{L_{\odot}}
 \label{eq11}
\end{equation}

The absolute magnitude, $M_{V}$, is related to the bolometric magnitude, $M_{bol}$, via:

\begin{equation}
 {M_{V}}={M_{bol}} - BC
 \label{eq12}
\end{equation}

Where $BC$ is the bolometric correction given by \cite{Reed+1998}:
\begin{equation}
BC=-8.499{\left [{\log {(T)}}-4\right ]^{4}}+13.421{\left[{\log {(T)}}-4\right ]^{3}}-8.131{\left [{\log{(T)}}-4\right ]^{2}} -3.901\left [{\log{(T)}}-4\right]-0.438
\label{eq13}
\end{equation}

We used the relation in equation~\ref{eq12} to calculate the absolute magnitude for all three systems. Using the interstellar extinction value ($A_{\nu}$) corresponding to the equatorial coordinates J2000 obtained from \cite {Schlafly+Finkbeiner+2011} for the $V$-band at effective wavelength =5517~\AA\ .
Finally we found that, for system KIC 2715417 ($A_{\nu}= 0.427$), for the System 6050116 ($A_{\nu}=0.369$) and for the system 6287172 ($A_{\nu}=0.339$). Using the distance modulus relation to calculate the distance of the three systems as follows:
\begin{equation}
 {D}={10^{0.2~(m-M+5-{A_{\nu}})}}
 \label{eq14}
\end{equation}

Where the distance is in parsec (pc), $m$ represents the apparent magnitude and $M_{V}$ is the absolute magnitude. All the absolute parameters (mass, semi major axis, radius, luminosity, bolometric magnitude and the distance) of the three systems which have been calculated are listed in Tables \ref{tab5}, \ref{tab6}, and \ref{tab7}.

\begin{table}[h]
\bc
\begin{minipage}[]{100mm}
\caption[]{\small {The absolute parameters of the system KIC 2715417\label{tab5}}}
\end{minipage}
\setlength{\tabcolsep}{2.5pt}
\small
 \begin{tabular}{lccc}
  \hline\noalign{\smallskip}
Parameter~~~~~~&~~~~~~Harmanec~~~~~~~~~~&~~~~~~~~~~Maceroni~~~~~~~~~~&~~~~~~~~~~Gazeas~~~~~~~~~~\\
  \hline\noalign{\smallskip}
$M_{1}(M_{\odot})$ & 0.902 $\pm$ 0.076 & 0.888 $\pm$ 0.032 & 0.877 $\pm$ 0.132   \\
$M_{2}(M_{\odot})$ & 0.261 $\pm$ 0.022 &  0.257 $\pm$ 0.009 & 0.254 $\pm$ 0.038   \\
$a (R_{\odot})$ & 1.692 $\pm$ 0.049 & 1.683 $\pm$ 0.020 & 1.677 $\pm$ 0.080  \\
$R_{1}(R_{\odot})$ & 0.781 $\pm$ 0.023 &  0.778 $\pm$ 0.009 & 0.774 $\pm$ 0.037   \\
$R_{2}(R_{\odot})$ & 0.497 $\pm$ 0.014 &  0.497 $\pm$ 0.014 & 0.492 $\pm$ 0.024   \\
$L_{1}(L_{\odot})$ & 0.399 $\pm$ 0.023 &  0.394 $\pm$ 0.009 & 0.391 $\pm$ 0.038  \\
$L_{2}(L_{\odot})$ & 0.124 $\pm$ 0.007 & 0.123 $\pm$ 0.003 & 0.122 $\pm$ 0.012 \\
$M_{bol1}$ & 5.739 $\pm$ 0.064 & 5.750 $\pm$ 0.026 & 5.759 $\pm$ 0.102   \\
$M_{bol2}$ & 7.002 $\pm$ 0.064 & 7.014 $\pm$ 0.026 & 7.023 $\pm$ 0.102  \\
$D (pc)$ & 324 $\pm$ 9  &  322 $\pm$ 4 & 320 $\pm$ 16 \\
  \noalign{\smallskip}\hline
\end{tabular}
\ec
\end{table}
 \FloatBarrier

\begin{table}[h]
\bc
\begin{minipage}[]{100mm}
\caption[]{\small {The absolute parameters of the system KIC 6050116\label{tab6}}}
\end{minipage}
\setlength{\tabcolsep}{2.5pt}
\small
 \begin{tabular}{lccc}
  \hline\noalign{\smallskip}
Parameter~~~~~~&~~~~~~Harmanec~~~~~~~~~~&~~~~~~~~~~Maceroni~~~~~~~~~~&~~~~~~~~~~Gazeas~~~~~~~~~~\\
  \hline\noalign{\smallskip}
$M_{1}(M_{\odot})$ & 0.702  $\pm$ 0.059 & 0.788  $\pm$ 0.063 & 0.887 $\pm$ 0.133   \\
$M_{2}(M_{\odot})$ & 0.402 $\pm$ 0.034 &  0.452 $\pm$ 0.036 & 0.508 $\pm$ 0.076   \\
$a (R_{\odot})$ & 1.680 $\pm$ 0.049 & 1.746 $\pm$ 0.048 & 1.816 $\pm$ 0.086  \\
$R_{1}(R_{\odot})$ & 0.757 $\pm$ 0.022 &  0.741 $\pm$ 0.020 & 0.771 $\pm$ 0.037   \\
$R_{2}(R_{\odot})$ & 0.606 $\pm$ 0.018 &  0.582 $\pm$ 0.016 & 0.605 $\pm$ 0.029   \\
$L_{1}(L_{\odot})$ & 0.225 $\pm$ 0.013 &  0.216 $\pm$ 0.012 & 0.233 $\pm$ 0.023  \\
$L_{2}(L_{\odot})$ & 0.131 $\pm$ 0.007 & 0.121 $\pm$ 0.007 & 0.131 $\pm$ 0.013 \\
$M_{bol1}$ & 6.360 $\pm$ 0.064 & 6.406 $\pm$ 0.060 & 6.320 $\pm$ 0.101   \\
$M_{bol2}$ & 6.945 $\pm$ 0.064 & 7.032 $\pm$ 0.060 & 6.946 $\pm$ 0.101  \\
$D (pc)$ & 233 $\pm$ 7  &  228 $\pm$ 6 & 237 $\pm$ 11 \\
  \noalign{\smallskip}\hline
\end{tabular}
\ec
\end{table}
 \FloatBarrier

\begin{table}[h]
\bc
\begin{minipage}[]{100mm}
\caption[]{\small {The absolute parameters of the system KIC 6287172\label{tab7}}}
\end{minipage}
\setlength{\tabcolsep}{2.5pt}
\small
 \begin{tabular}{lccc}
  \hline\noalign{\smallskip}
Parameter~~~~~~&~~~~~~Harmanec~~~~~~~~~~&~~~~~~~~~~Maceroni~~~~~~~~~~\\
  \hline\noalign{\smallskip}
$M_{1}(M_{\odot})$ & 1.376 $\pm$ 0.12 & 1.239 $\pm$ 0.054   \\
$M_{2}(M_{\odot})$ & 0.883 $\pm$ 0.070 &  0.751 $\pm$ 0.033   \\
$a (R_{\odot})$ & 1.899 $\pm$ 0.055 & 1.834 $\pm$ 0.026   \\
$R_{1}(R_{\odot})$ & 0.934 $\pm$ 0.027 &  0.902 $\pm$ 0.013  \\
$R_{2}(R_{\odot})$ & 0.779 $\pm$ 0.023 &  0.752 $\pm$ 0.011    \\
$L_{1}(L_{\odot})$ & 1.532 $\pm$ 0.087 &  1.429 $\pm$ 0.041 \\
$L_{2}(L_{\odot})$ & 1.049 $\pm$ 0.060 & 0.978 $\pm$ 0.028  \\
$M_{bol1}$ & 4.277 $\pm$ 0.064 & 4.352 $\pm$ 0.031    \\
$M_{bol2}$ & 4.688 $\pm$ 0.064 & 4.764 $\pm$ 0.031   \\
$D (pc)$ & 401 $\pm$ 12  &  387 $\pm$ 6  \\
  \noalign{\smallskip}\hline
\end{tabular}
\ec
\tablecomments{0.86\textwidth}{The system KIC 6287172 belongs to ELV type. Gazeas \& Niarchos (2006) Method is not included in Table 7 because it is only applicable to W UMa-type.}
\end{table}
 \FloatBarrier

\section {EVOLUTIONARY STATE OF THE SYSTEMS}

\begin{figure}[h] 
  \centering
 \includegraphics[width=8.0cm, angle=0]{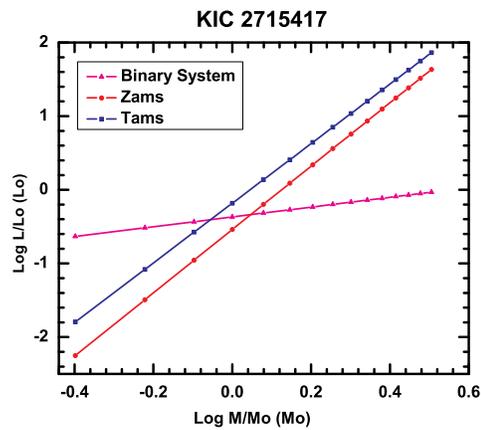}
 \begin{minipage}[]{85mm}
\caption{L-M of Binary System KIC 2715417\label{Fig19}. } 
\end{minipage}
\end{figure}

\begin{figure}[h] 
  \centering
 \includegraphics[width=8.0cm, angle=0]{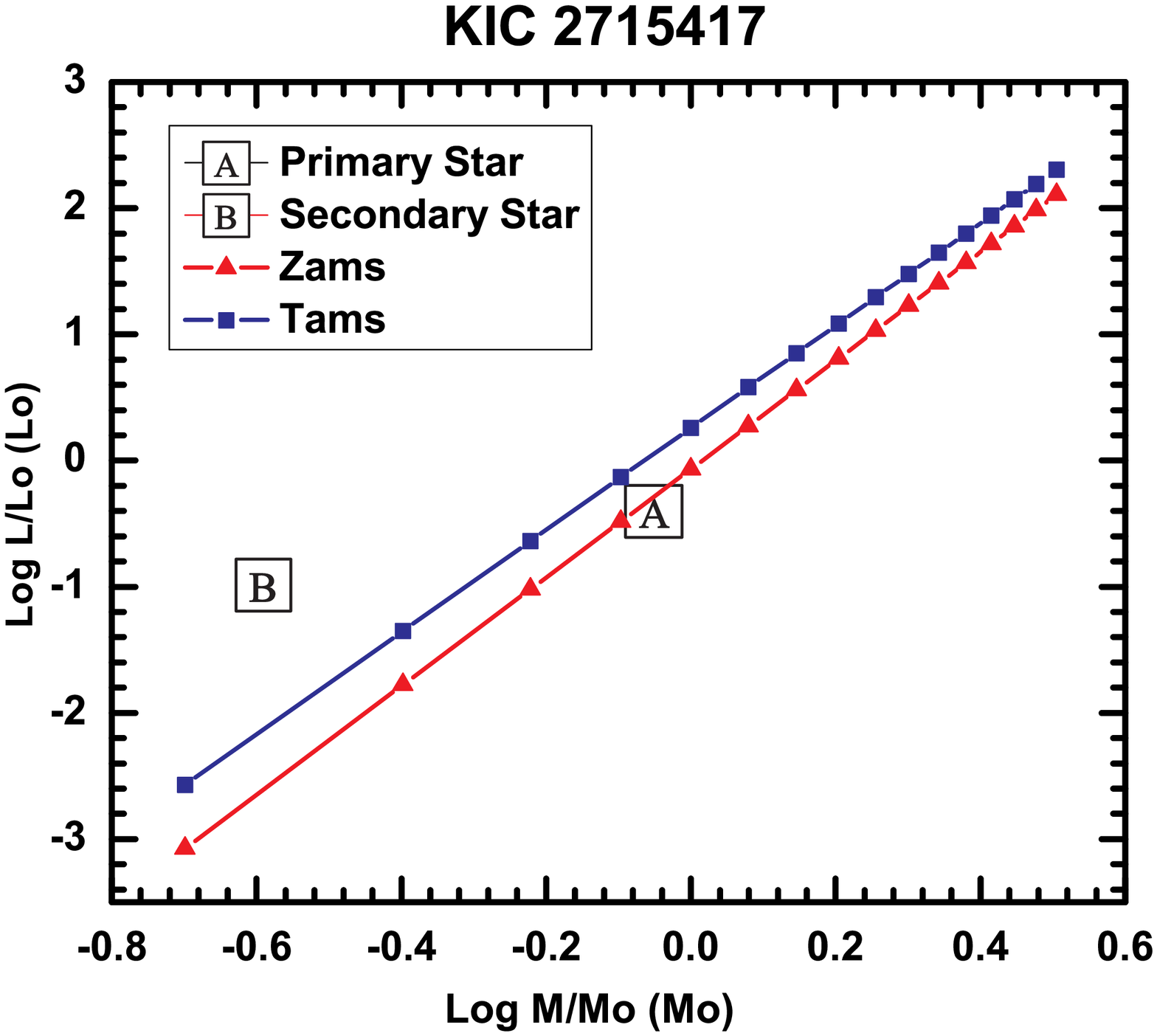}
 \begin{minipage}[]{85mm}
\caption{L-M relation of the system KIC 2715417\label{Fig20}. } 
\end{minipage}
\end{figure}

\begin{figure}[h] 
  \centering
 \includegraphics[width=8.0cm, angle=0]{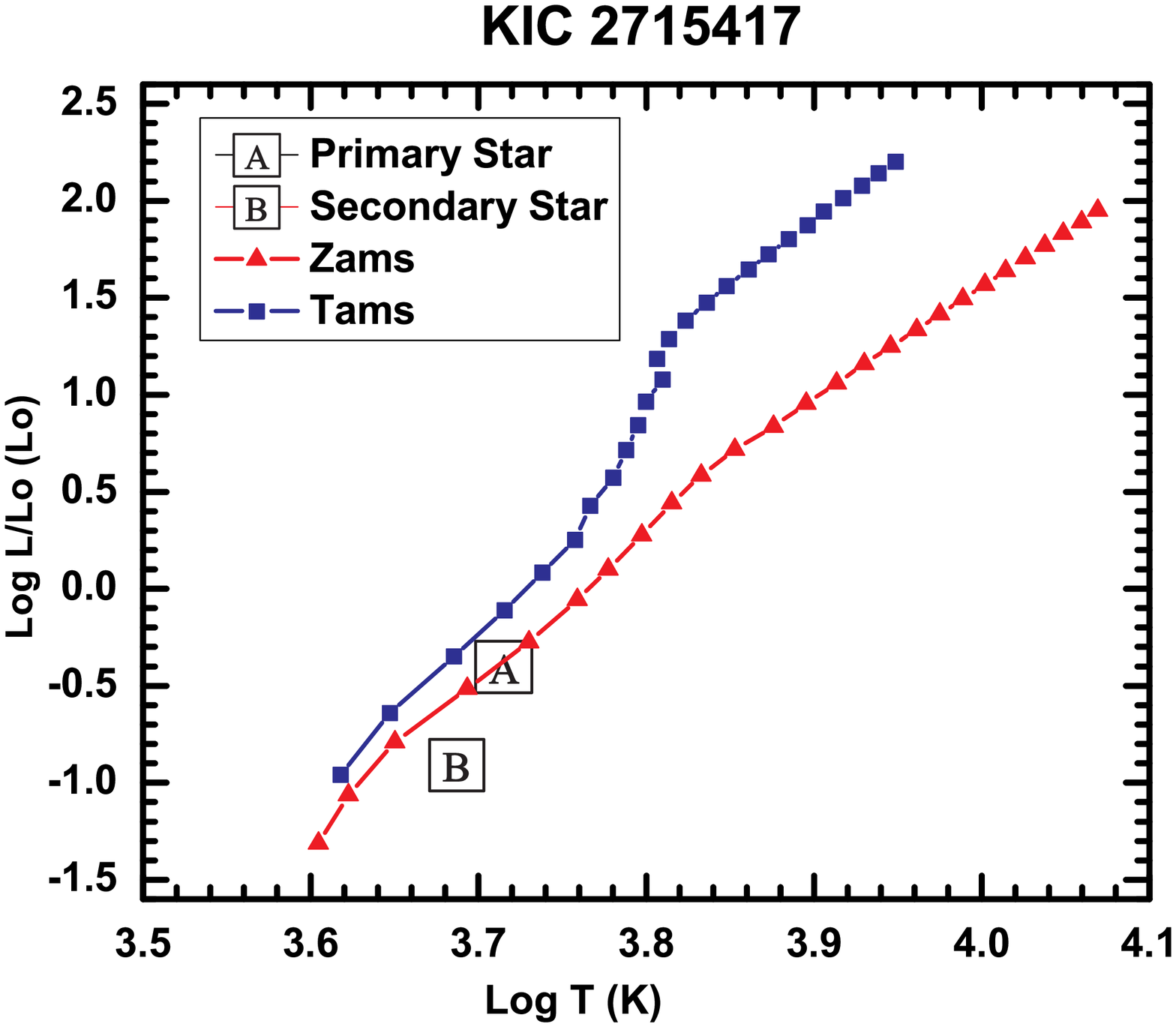}
 \begin{minipage}[]{85mm}
\caption{L-T relation of the system KIC 2715417\label{Fig21}. } 
\end{minipage}
\end{figure}

\begin{figure}[h] 
  \centering
 \includegraphics[width=8.0cm, angle=0]{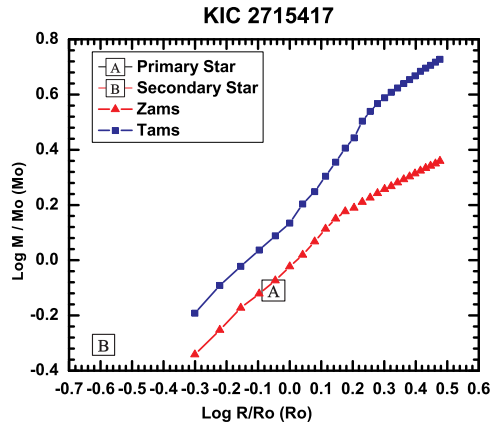}
 \begin{minipage}[]{85mm}
\caption{M-R relation of the system KIC 2715417\label{Fig22}. } 
\end{minipage}
\end{figure}

\begin{figure}[h] 
  \centering
 \includegraphics[width=8.0cm, angle=0]{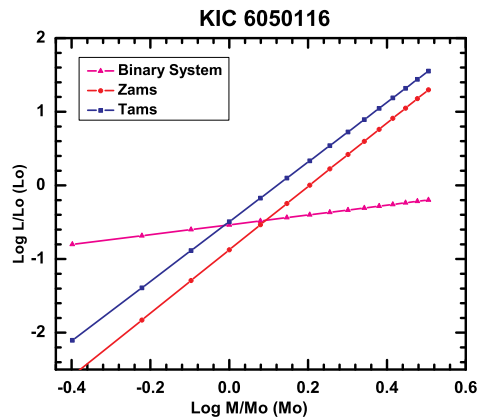}
 \begin{minipage}[]{85mm}
\caption{L-M of Binary system KIC 6050116\label{Fig23}.} 
\end{minipage}
\end{figure}

\begin{figure}[h] 
  \centering
 \includegraphics[width=8.0cm, angle=0]{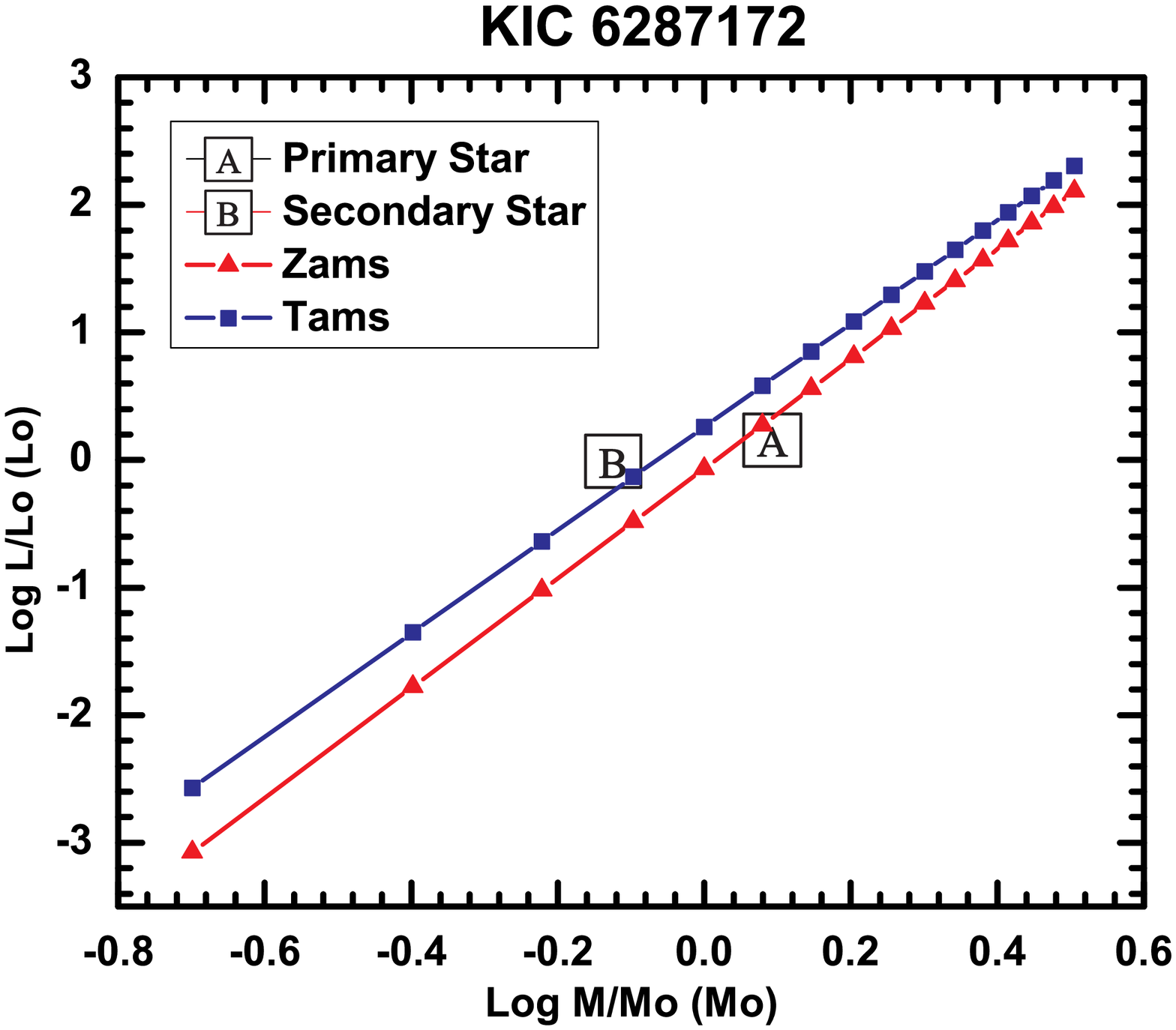}
 \begin{minipage}[]{85mm}
\caption{L-M relation of the system KIC 6050116\label{Fig24}. } 
\end{minipage}
\end{figure}

\begin{figure}[h] 
  \centering
 \includegraphics[width=8.0cm, angle=0]{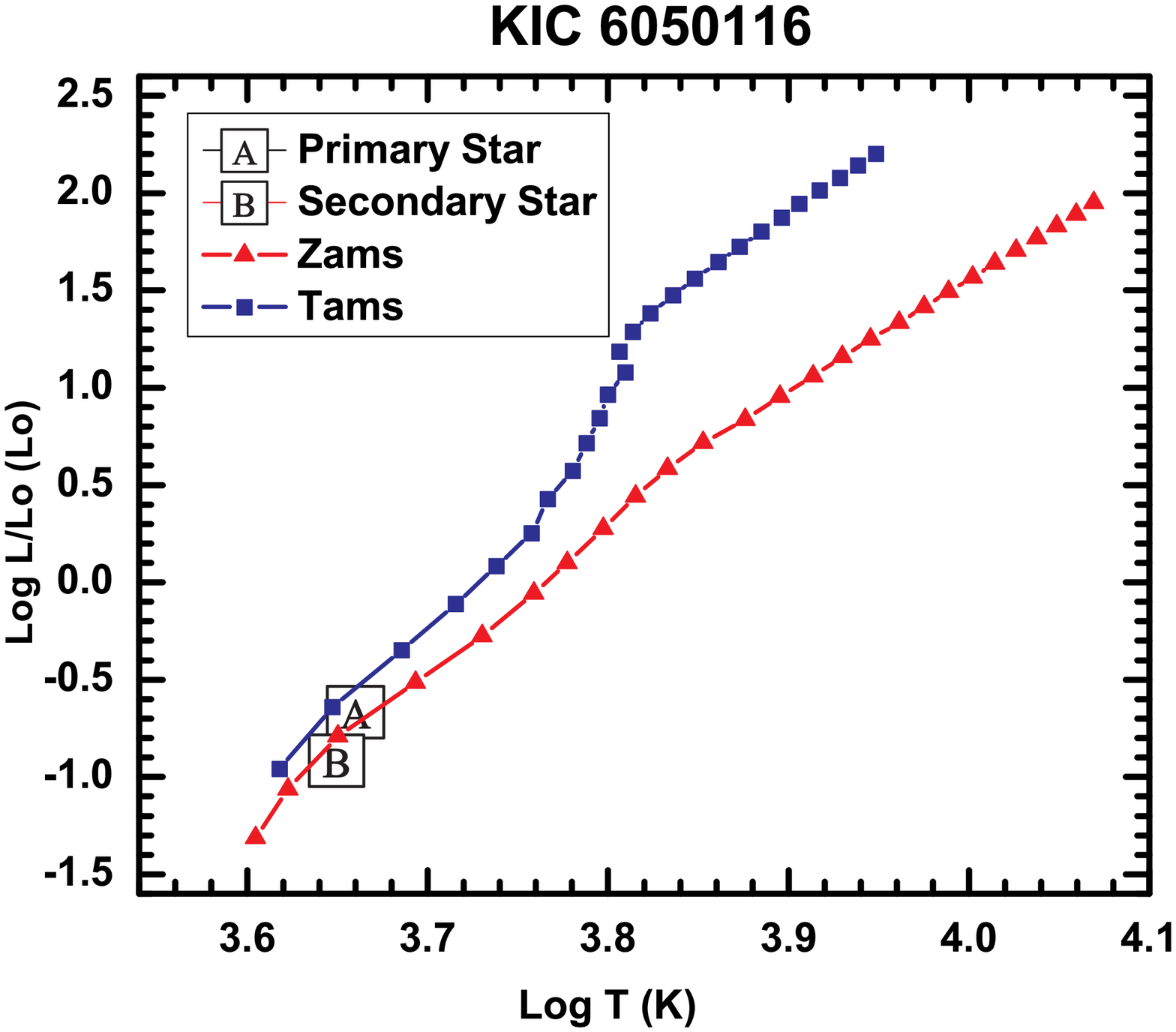}
 \begin{minipage}[]{85mm}
\caption{L-T relation of the system KIC 6050116\label{Fig25}. } 
\end{minipage}
\end{figure}

\begin{figure}[h] 
  \centering
 \includegraphics[width=8.0cm, angle=0]{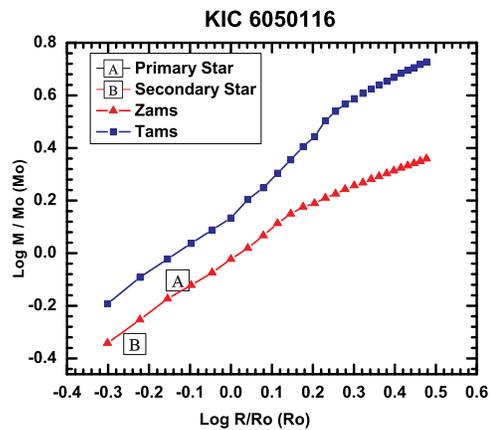}
 \begin{minipage}[]{85mm}
\caption{M-R relation of the system KIC 6050116\label{Fig26}. } 
\end{minipage}
\end{figure}

\begin{figure}[h] 
  \centering
 \includegraphics[width=8.0cm, angle=0]{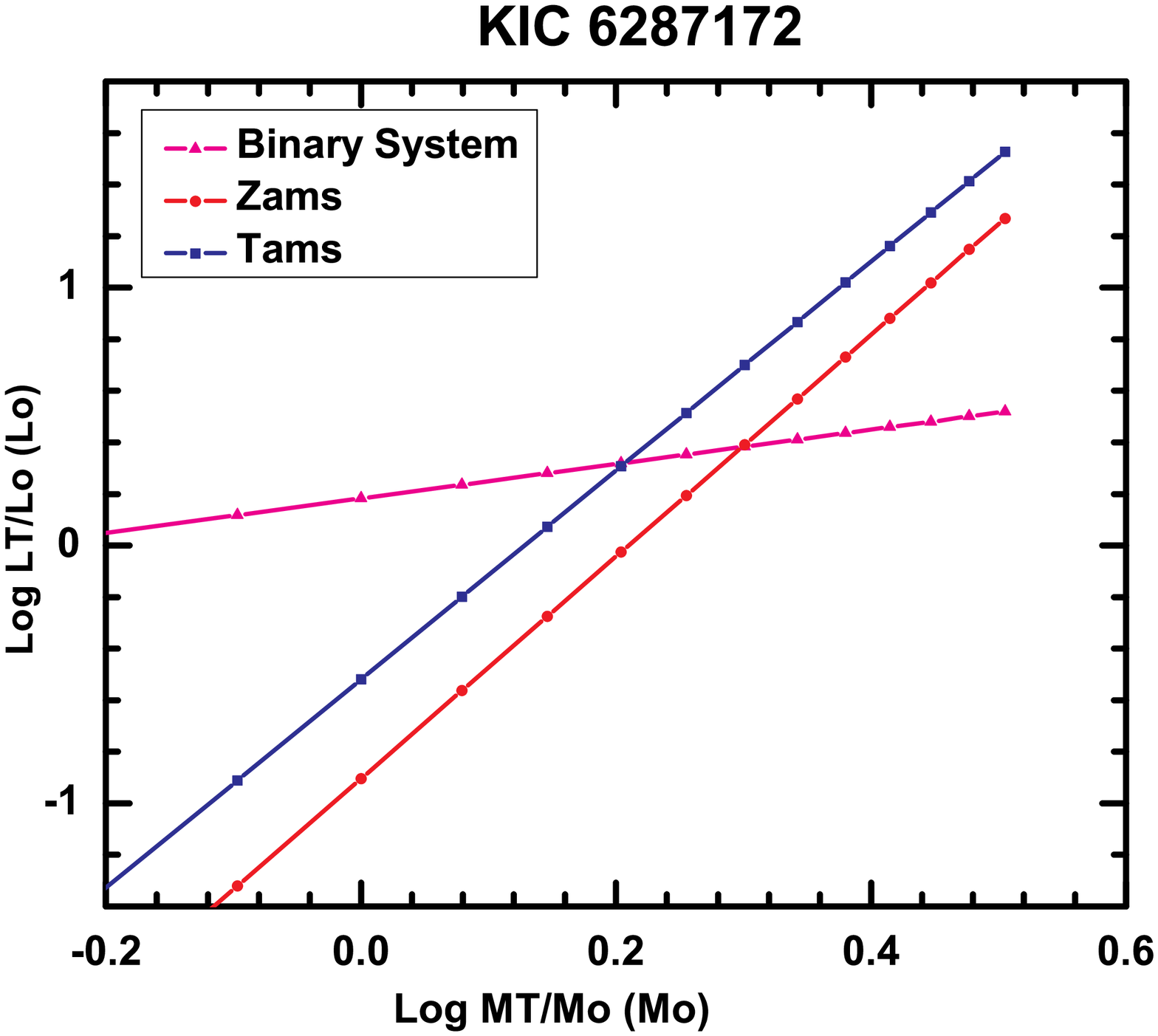}
 \begin{minipage}[]{85mm}
\caption{L-M of Binary system KIC 6287172\label{Fig27}. } 
\end{minipage}
\end{figure}

\begin{figure}[h] 
  \centering
 \includegraphics[width=8.0cm, angle=0]{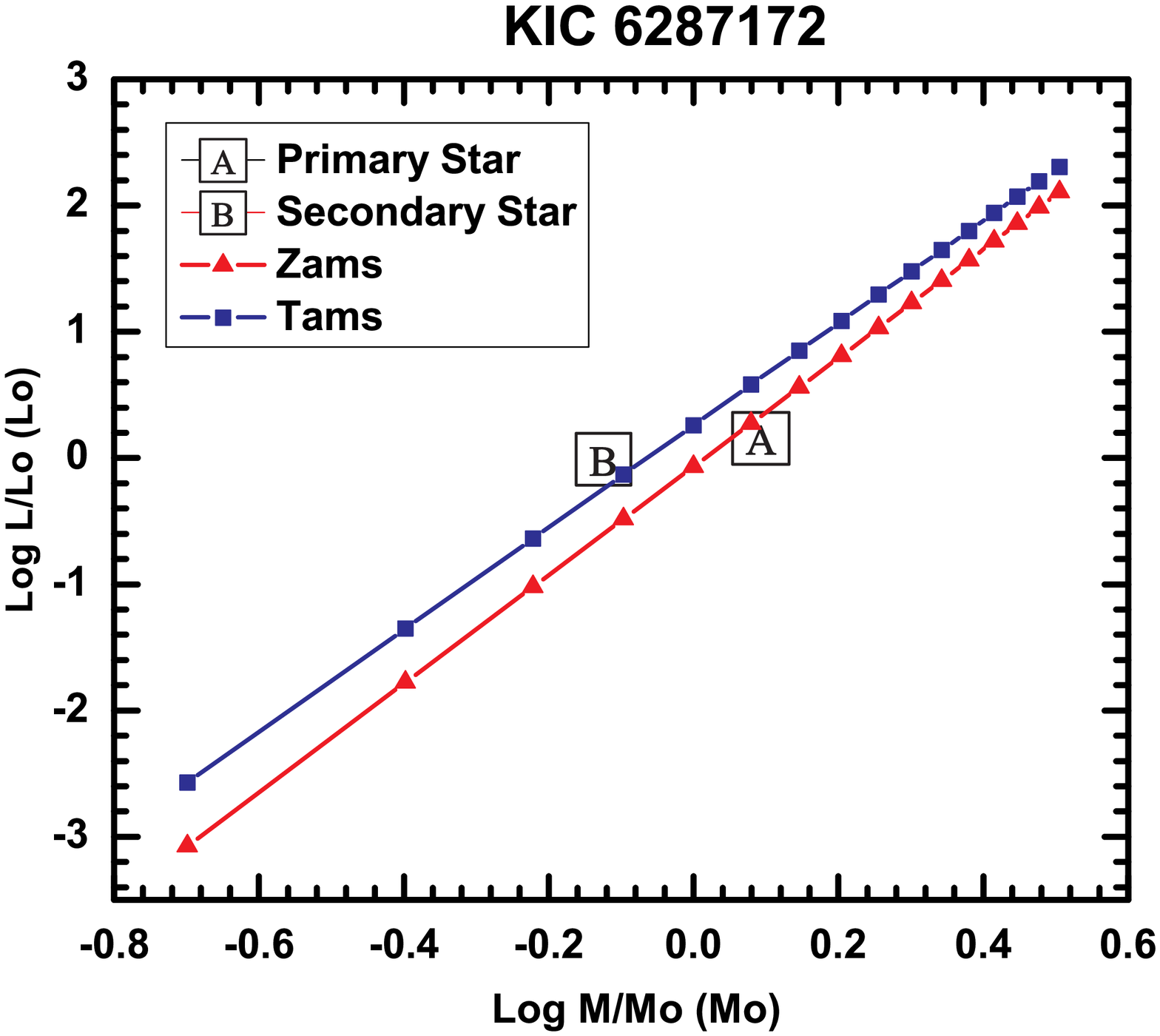}
 \begin{minipage}[]{85mm}
\caption{L-M relation of the system KIC 6287172\label{Fig28}. } 
\end{minipage}
\end{figure}

\begin{figure}[h] 
  \centering
 \includegraphics[width=8.0cm, angle=0]{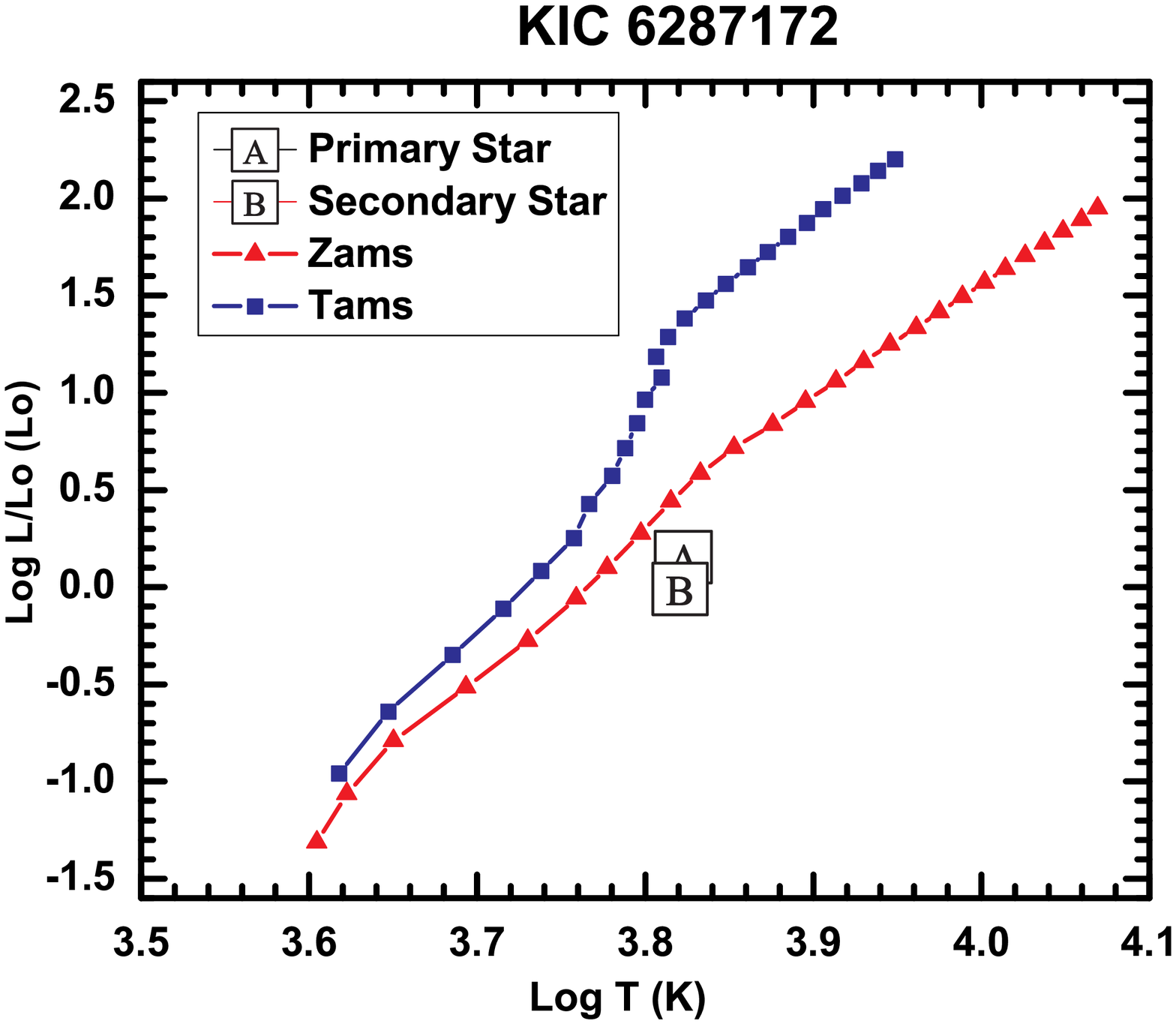}
 \begin{minipage}[]{85mm}
\caption{L-T relation of the system KIC 6287172\label{Fig29}. } 
\end{minipage}
\end{figure}

\begin{figure}[h] 
  \centering
 \includegraphics[width=8.0cm, angle=0]{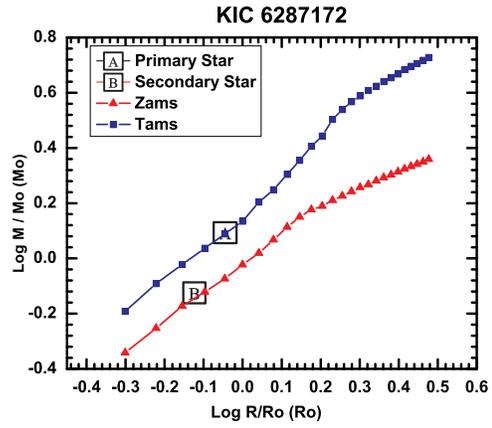}
 \begin{minipage}[]{85mm}
\caption{M-R relation of the system KIC 6287172\label{Fig30}. } 
\end{minipage}
\end{figure}

In order to study the evolutionary state of the three systems (KIC 2715417, KIC 6050116, and KIC 6287172), we have plotted the physical parameters listed in Tables  \ref{tab5}, \ref{tab6} and \ref{tab7}  of the components of the three systems on the $H-R$, $M-R$ and $M-L$ diagrams in Figures \ref{Fig19} -\ref{Fig30}. Using the evolutionary tracks for non-rotating model which have been computed by \cite{Mowlavi+etal+2012} for both zero age main sequence stars $(ZAMS)$ and terminal age main sequence stars $(TAMS)$ with metallicity $Z=0.014$ (Solar metallicity).
      As it is clear from Figures \ref{Fig19}, \ref{Fig23} and  \ref{Fig27} the straight line (Binary System) represents the relation between the total luminosity and the total mass of the binary system as written in equation~\ref{eq5}. We can determine the total mass from the intersection between the two lines, Binary System and $ZAMS$. In Figures \ref{Fig20}, \ref{Fig24} and  \ref{Fig28} the primary components of the three systems are located on $ZAMS$, however the secondary components of the three systems are located above $ZAMS$. This can be attributed to the rise of temperature of the secondary component due to thermal contact between the two components of the binary systems and consequently the value of luminosity will rise without any change in mass.

      In Figures \ref{Fig21}, \ref{Fig25} and \ref{Fig29} the primary components of the three systems are located on $ZAMS$, but for the secondary components of the three systems are located below $ZAMS$.  That may be due to thermal contact. In the figures \ref{Fig22} and \ref{Fig26} the primary components of the two systems (KIC 2715417, KIC 6050116) are located on $ZAMS$, but the secondary component of these two systems are located faraway from $ZAMS$.  In Figure \ref{Fig30} we note that the primary component of the system KIC 6287172 is located above $ZAMS$ with additional mass accumulated due to matter transfer from the secondary component.
  This system belongs to the type of Ellipsoidal Variables $(ELV)$. The type of $ELV$ can be described as very close non-eclipsing binaries whose two components are of non-spherical shape due to the mutual gravitation. Their light variation can change as seen from the Earth as a result of the rotation of the two components where their surfaces facing the observers are changing.

\subsection{KIC 2715417}
 \FloatBarrier
 
\subsection{KIC 6050116}
 \FloatBarrier
 
\subsection{KIC 6287172}
 \FloatBarrier
 
\section{CONCLUSIONS AND DISCUSSIONS}

We presented the physical properties of the systems KIC 2715417, KIC 6050116 and 
KIC 6287172 derived from the analyses of the Kepler data according to the above results, we arrived to following conclusions: 
  The system KIC 2715417 is considered a near contact system. The second system KIC 6050116 is an overcontact system, while the system KIC 6287172 belongs to ellipsoidal variable $(ELV)$ or non-eclipsing overcontact type. The effective temperature of the primary star for all the three systems are slightly higher than the secondary component. We found that the primary components for our systems are located on $ZAMS$ line while the secondary components are located away from $ZAMS$. This can be due to the heating of the secondary components of the close binaries by thermal contact.
  
    We used the three mentioned methods to determine the masses of the primary components of the binary systems and calculated the masses of the secondary components from the photometric mass ratios obtained from the $Q$-search method which gives good values compared with the spectroscopic mass ratio particularly for contact type. This enabled us to obtain the required comparison of the systems under considerations. It is found that the method of \cite{Maceroni+Van'tVeer+1996} gives the best results for our close binary systems which have thermal contact between the two components.
    
    For the system KIC 2715417 the primary component has a mass of $0.888 \pm 0.032 M_{\odot}$ while the secondary component has a mass $0.257~\pm~0.009 M_{\odot}$. The distance equal $322~\pm~4 Pc$. Each component has a cool spot while the spectral type is $K0$ and $K2$ for the primary and secondary components respectively. The system KIC 6050116 consists of two components with masses $0.788~\pm~0.063 M_{\odot}$ and $0.452~\pm~0.036 M_{\odot}$ for the primary and secondary components respectively. This system is located at a distance $228~\pm~6~Pc$. Only the primary component has a cool spot, while the spectral type is $K_{5}$ for both two components of the system. 
    
    The third binary system KIC 6287172 has larger masses $1.239~\pm~0.054 M_{\odot}$ and $0.751~\pm~0.033~M_{\odot}$ for the primary and secondary components respectively. The system is further away than the previous two systems at a distance of $387~\pm~6~Pc$, while the spectral type is $F_{5}$ for both primary and secondary components without any spot for both components. The difference between two maxima in any light curve indicates the presence of spot on the surface of the component which is called O'Connell effect \citep{O'Connell+1951}. The presence of cool spots on the primary and secondary components of the KIC 2715417 and the primary component of the system KIC 6050116 indicates the existence of magnetic field inside the spots which inhibit convection similar to the case of sunspots.
    
   \normalem
\begin{acknowledgements}
       
    In this paper we collected all the data from Kepler mission. Kepler was selected as the 10th mission of the Discovery Program. Also we used the Vizier database maintained at CDS, Strasbourg, France. This work is part of M.Sc. thesis submit to Cairo University by one of us (M. A. NegmEldin).

\end{acknowledgements}
  
\bibliographystyle{raa}
\bibliography{bibtex}

{\bfseries {References}}\\

Deb, S., Singh, H. P., Seshadri, T. R., \& Gupta, R. 2010, NewA, 15, 662\\
Drilling, J. S., \& Landolt, A. U. 2002, Normal Stars. In: Cox A.N. (eds) Allen's Astrophysical Quantities.
Springer, New York, NY, 381\\
Gazeas, K. D., \& Niarchos, P. G. 2006, MNRAS, 370, L29\\
Gazeas, K., \& Stepien, K. 2008, MNRAS, 390, 1577\\
Harmanec, P. 1998, BAICz, 39, 329\\
Lucy, L. B. 1967, ZA, 65, 89\\
Lucy, L. B., \& Wilson, R. E. 1979, APJ, 231, 502\\
Maceroni, C., \& Van'tVeer, F. 1996, A\&A, 311, 523\\
Mowlavi, N., Eggenberger, P., Meynet, G., et al. 2012, A\&A, 541, A41\\
O'Connell, D. J. 1951, PRCO, 2, 85\\
Popper, D. M. 1980, ARA\&A, 18, 115\\
Prsa, A., \& Zwitter, T. 2005, ApJ, 628, 426\\
Reed, B. C. 1998, JRASC, 92, 36\\
Ruciski, S. M. 1969, AcA, 19, 245\\
Schlafly, E. F., \& Finkbeiner, D. P. 2011, ApJ, 737, 103\\
VanHamme, W. 1993, AJ, 106, 2096\\
Wilson, R. E., \& Devinney, E. J. 1971, ApJ, 166, 605\\

\end{document}